\documentclass[12pt]{article}
\usepackage{amsfonts}
\usepackage{amsmath}
\usepackage{latexsym,amssymb,enumerate,ulem,pifont,calc}
\usepackage{graphicx}
\usepackage{caption}
\usepackage{subcaption}
\usepackage[utf8]{inputenc}
\usepackage[toc,page]{appendix}
\usepackage{amsmath}
\usepackage{gensymb}
\usepackage{multicol}
\usepackage{a4wide,graphics,amsmath,amssymb,cite,nicefrac,mathrsfs}
\usepackage[verbose]{wrapfig}
\usepackage{authblk,hyperref}
\usepackage{url,amsfonts}
\usepackage{dirtytalk}
\usepackage{xcolor}

  \numberwithin{equation}{section}

\begin{document}

\begin{titlepage}

		\vskip 1.0cm

	\begin{center}
			{\boldmath \Huge{ Stable Non-BPS Black Holes and \\ {~} \\  Black Strings in Five Dimensions}}

		\vskip 2.0cm
		{\bf \large Anshul Mishra \footnote{
E-mail - anshulmishra2025@gmail.com} and \large Prasanta K. Tripathy \footnote{E-mail - prasanta@iitm.ac.in}  }

			\vspace*{1cm}
				\vskip 0.5cm
			{\it Centre for Strings, Gravitation and Cosmology, \\
				Department of Physics, \\
				Indian Institute of
				Technology
				Madras, Chennai
				600 036, India}\\
			\vspace{0.2cm}

		\vskip 10pt
			\end{center}
		
		\vskip 1.5cm
		
		\begin{center}
			
			 {\bf ABSTRACT}
		
			 \end{center}%
			 
		In this paper we study black hole and black string solutions in five dimensional $N=2$ 
		supergravity theories arising from the compactification of M-theory on Calabi-Yau
		manifolds. In particular, we consider explicit examples of three parameter Calabi-Yau manifolds 
		which are obtained as hypersurfaces in toric varieties. Using the attractor mechanism, 
		we obtain BPS as well as non-BPS black holes in these compactified supergravity 
		theories. We also consider the black string solutions in these models. We analyze 
		the stability of these extremal black brane configurations by computing the recombination factor.
		We find multiple stable non-BPS attractor solutions in some of these models.
			 
	\end{titlepage}

\section{Introduction}

It is well known that the compactification of string theory gives rise to a landscape of 
consistent low energy effective theories in various dimensions \cite{Douglas:2003um}. This landscape has further
been enriched by the swampland program \cite{Ooguri:2006in,Vafa:2005ui} which contrived a set of general criteria that the 
low energy effective theories must hold. One of the key ingredients devised in order to 
distinguish the low energy effective theories among each other is the weak gravity conjecture \cite{Arkani-Hamed:2006emk}.
This conjecture tells that gravity is the weakest force among all and as a consequence objects
with large charge must decay into their respective elementary constituents unless they are 
protected by a symmetry. 

In gravity theories, black holes provide simplest objects which can be used to further 
analyze the weak gravity conjecture. Of particular interest are black hole configurations
in $N=2$ supergravity theories. For these theories, the attractor mechanism  \cite{Ferrara:1995ih, Ferrara:1996dd, Ferrara:1996um, Ferrara:1997tw, Goldstein:2005hq} provides a 
suitable technique to construct the extremal black hole solutions. These extremal black
holes may or may not preserve supersymmetry \cite{Ferrara:1997tw,Goldstein:2005hq}. The attractor mechanism ensures that the 
scalar fields in these theories, which in general may take arbitrary values at spatial 
infinity, must run into fixed points at the horizon of the corresponding extremal black 
hole. Their values at the horizon are completely determined in terms of the black hole 
charges. The attractor points in the moduli space correspond to critical points of a 
suitably constructed black hole effective potential \cite{Ferrara:1997tw}. This in turn determines the mass 
as well as entropy of these black holes.

Recently Long, Sheshmani, Vafa and Yau have constructed a number of non-BPS black branes 
using the attractor mechanism \cite{Long:2021lon}. They have considered $N=2$ supergravity theories in five dimensions \cite{Gunaydin:1984ak}, obtained 
upon the compactification of M-theory on Calabi-Yau manifolds \cite{Cadavid:1995bk, Ferrara:1996hh}. $M2$ branes wrapped on two
cycles of the Calabi-Yau manifold give rise to black holes and $M5$ branes wrapped on 
four cycles give rise to black string configurations. Depending upon whether the two 
cycle corresponds to a holomorphic or a non-holomorphic curve one obtains a BPS or a 
non-BPS black hole. Likewise, if the four cycle is a (non)-holomorphic divisor one 
obtains a (non)-BPS black string. 

BPS objects, being protected by supersymmetry remain stable. Whereas the same thing 
may not hold for non-BPS configurations and a separate study is needed to address the issues of their 
stability. For non-BPS black holes, the attractor mechanism determines the black 
hole entropy. This, in turn is conjectured \cite{Long:2021lon} to give rise to the asymptotic volume of the 
non-holomorphic curve on which the $M2$ brane is wrapped provided the non-BPS black 
hole is doubly extremal. Similarly, for doubly extremal non-BPS black strings, the 
string tension determines the asymptotic volume of the non-holomorphic divisor on which 
the corresponding $M4$ brane is wrapped. These non-holomorphic cycles correspond
to local volume minimizers in a given homology class. If, within the same homology
class there exists a piece-wise calibrated representative with a smaller volume then
the black brane decays into the constituent brane/anti-brane pairs wrapping this
piece-wise representative cycle in accordance of the weak gravity conjecture. On the 
other hand, if no such piece-wise calibrated representative with a smaller volume exists 
within the same homology class then the black brane becomes stable against decay into 
constituent brane-anti-brane configurations. In the later case, the constituent brane 
configurations are said to be undergoing a recombination in order to form a stable 
non-BPS black brane \cite{Long:2021lon}. Several examples of Calabi-Yau manifolds has been considered and 
it has been observed that the non-BPS black holes are always unstable \cite{Long:2021lon}. On the other hand, 
it has been observed that some of these models admit stable non-BPS black strings \cite{Long:2021lon}.

These results have further been extended in \cite{Marrani:2022jpt} to consider an exhaustive study of all the
two parameter Calabi-Yau compactifications of M-theory. It has been observed that the 
resulting non-BPS black hole configurations in all these examples are unstable and 
decay into the constituent BPS/non-BPS pairs. Moreover, for a given charge configuration, 
there exists a unique non-BPS black hole attractor. In contrast, there exist several
non-BPS black string configurations which are stable against decay into constituent
BPS/non-BPS pairs. In addition, many of these models also admit multiple basin of 
non-BPS black string attractors for a given charge configuration. Similar phenomena
have also been observed in \cite{Aspman:2022vlx}, in the context of four dimensional $N=2$ supergravity theories 
arising from the Calabi-Yau compactifications of type-$IIA$ string theory. The authors 
have considered non-BPS $D0-D4$ as well as $D0-D2-D4$ black hole configurations in one 
and two parameter Calabi-Yau models and made similar analysis. While many of these 
configurations are unstable, there are also examples of several stable non-BPS black 
holes in these classes of compactifications.

The authors in \cite{Long:2021lon} considered a few examples of Calabi-Yau models with small $h_{11}$
and interestingly  the non-BPS black holes obtained in all their examples remained 
unstable. Surprisingly, this result holds true in all the two parameter examples 
considered in \cite{Marrani:2022jpt}. It remains to be seen whether this result holds for all five
dimensional non-BPS black holes resulting in Calabi-Yau compactifications of M-theory.
Further, the analysis of \cite{Marrani:2022jpt} shows the existence of multiple basin of attractors in
some of these two-parameter models. In models with larger $h_{11}$ there is a possibility 
of obtaining more interesting black hole and black string configurations. In the present 
work we wish to generalize these results for a number of three parameter Calabi-Yau 
manifolds obtained as hypersurfaces in toric varieties. The cohomology data for these 
manifolds have been computed. Using these data we can obtain the black brane configurations
and study their properties. In this paper we consider some of these three parameter
models to obtain extremal black branes in them and analyze their properties.

The plan of the paper is as follows. First we will review the basic results on extremal 
black hole and black string attractors in five-dimensional minimal supergravity in  
\S\ref{sec:Review of 5D Attractor}. Also in this section we will set up our notations 
and conventions and outline the attractor equations that needs to be analyzed further.
In \S\ref{sec:Three Parameter Model}, we will shift our focus towards three parameter 
Calabi-Yau manifolds where we will outline some useful formulae and introduce some 
notations that will be used throughout the paper. Following a comprehensive treatment of BPS black holes in 
\S\ref{BPS Black holes} we turn our attention in particular to the black hole solutions
resulting from the Calabi-Yau manifolds emerging as hypersurfaces in toric varieties
(THCY). Subsequently, our investigation extends to non-BPS black hole solutions in 
\S\ref{sec:Non BPS Black Holes}. In \S\ref{BHcase1} we take different examples
of three parameter THCY models and investigate non-BPS black hole configurations. 
We also show the existence of  multiple non-BPS black hole solutions.
In \S\ref{stbl1} we compute the recombination factor for non-BPS extremal black 
hole attractors in these THCY models and show that there exists stable non-BPS 
black holes. Moving forward, in \S\ref{Non Bps black strings} we outline the essential 
details concerning extremal black string attractors in five-dimensional, minimal supergravity. Once again, 
we turn our attention into three-parameter Calabi-Yau manifolds. We take different cases 
for non-BPS black strings and obtain multiple solutions. We compute the 
recombination factor corresponding to these non-BPS extremal black strings in \S\ref{stbl2} highlighting the 
stability of these solutions. Finally we summarize our findings in \S\ref{cncl}.

\section{Review of 5D Black Hole Attractor} \label{sec:Review of 5D Attractor}

In this section we will review the basics of black brane attractors in $N=2$ supergravity 
theory in five dimensions. We consider the five dimensional low energy effective theory arising
upon the compactification of M-theory on a Calabi-Yau manifold. The corresponding Lagrangian 
density is given by\cite{Gunaydin:1983bi}
\begin{equation}
\frac{\mathcal{L}}{\sqrt{-\mathbf{g}}}=-\frac{1}{2}R-\frac{1}{4}G_{IJ}F_{\mu
\nu }^{I}F^{J|\mu \nu }-\frac{1}{2}g_{ij}\partial _{\mu }\varphi
^{i}\partial ^{\mu }\varphi ^{j}+\frac{1}{6^{3/2}\sqrt{-\mathbf{g}}}%
C_{IJK}\varepsilon ^{\lambda \mu \nu \rho \sigma }F_{\lambda \mu }^{I}F_{\nu
\rho }^{J}A_{\sigma }^{K} \ .
\end{equation}%
Here $\mathbf{g}_{\mu \nu }$
is the space-time metric and  $\mathbf{g}$ is its determinant. $R$ is the corresponding Ricci
scalar, $\varphi^i$ are the real scalar fields arising from the K\"ahler moduli $t^I  
(I=1,\cdots,h_{1,1}),$ and $A_\mu^I$ are the gauge fields in the vector multiplet with 
corresponding field strengths $F^I_{\mu\nu}$. The triple intersection numbers of the 
Calabi-Yau manifold are denoted by $C_{IJK}$. The volume of the Calabi-Yau manifold is 
constrained to satisfy
\begin{equation}\label{veq1}
C_{IJK} t^It^Jt^K = 1 \ . 
\end{equation}
Thus, only $h_{1,1}-1$ real scalar fields are independent and are denoted by $\phi^i$. The  
moduli space metric $g_{ij}$ is given by
\begin{equation}
g_{ij}=\frac{3}{2}\partial _{i}t^{I}\partial
_{j}t^{J}G_{IJ},~~g^{ij}g_{jk}=\delta _{k}^{i},
\end{equation}%
Here $G_{IJ}$ is the pull back of moduli space metric onto the ``ambient space" formed by
the K\"ahler moduli $t^I$. The metric $G_{IJ}$ is derived from the prepotential ${\cal V}
= C_{IJK} t^It^Jt^K$ as 
\begin{equation}
G_{IJ}:=-\frac{1}{3}\left. \frac{\partial ^{2}\log C_{LMN}t^{L}t^{M}t^{N}}{%
\partial t^{I}\partial t^{J}}\right\vert _{\ast }=\left(
3C_{ILM}C_{JNP}t^{L}t^{M}t^{N}t^{P}-2C_{IJM}t^{M}\right)\vert_{\ast },
\end{equation}%
where \textquotedblleft $\ast $" indicates that we need to impose the constraint \eqref{veq1}
after taking the derivatives.

The critical points of this theory have been studied extensively in \cite{Chou:1997ba} using
the attractor mechanism. The model admits extremal black hole as well as extremal black 
string solutions. We will first review the black hole configurations. These are electrically
charged objects carrying charges $q_I$. They are described in terms the black hole effective 
potential \cite{Chou:1997ba}
\begin{equation}
V=G^{IJ}q_{I}q_{J}=Z^{2}+\frac{3}{2}g^{ij}\partial _{i}Z\partial _{j}Z \ .
\label{V}
\end{equation}%
Here $Z$ denotes the central charge of the black hole $Z = q_It^I$. The critical points of this potential are given by 
\begin{equation}
\partial_iV = 0 \ .
\end{equation}
For a comprehensive treatment of the critical points using the so called new attractor
approach see \cite{Ferrara:2006xx,Cerchiai:2010xv}.
The above condition is obviously satisfied by the extrema of the central charge $Z$. All 
such critical points correspond to supersymmetry preserving extremal black holes. However,
there exist critical points which do not extremize $Z$. These correspond to non-supersymmetric
extremal black holes. For both these cases, the entropy of the black hole is determined by
the value $V_0$ of the effective potential \eqref{V} at the critical point. In the present 
convention, the black hole entropy is given by
\begin{equation}
S=2\pi \left( \frac{V_0}{9}\right) ^{3/4} \ . 
\end{equation}%

To obtain the equation of motion it is more convenient to express the black hole effective
potential as \cite{Marrani:2022jpt}
\begin{equation}
V=\frac{3}{2}Z^{2}-\frac{1}{2}C^{IJ}q_{I}q_{J}  \ . \label{V3}
\end{equation}%
Here we have introduced the notation $C_{IJ} = C_{IJK} t^K$ and $C^{IJ}$ is the matrix 
inverse of $C_{IJ}$. Upon extremizing the potential \eqref{V3} with the constraint 
\eqref{veq1}, we obtain 
\begin{equation}
3Z\big(q_{K}-ZC_{KIJ}t^{I}t^{J}\big)+\frac{1}{2}\big(%
C^{IL}C^{JM}C_{KLM}-C^{IJ}C_{KLM}t^{L}t^{M}\big)q_{I}q_{J}=0  \ . \label{5deom}
\end{equation}%
The supersymmetric critical points $\partial_i Z=0$ are given by
\begin{equation}
q_{K}-ZC_{KIJ}t^{I}t^{J}=0\ .  \label{bpseq}
\end{equation}%
Clearly, the first term of \eqref{5deom} vanishes if the above is satisfied. With a 
little bit of work it can be shown that the second term of this equation also vanishes
for the supersymmetric critical points. The equation of motion \eqref{5deom} also 
admits critical points for which \eqref{bpseq} does not hold. These correspond to 
non-BPS critical points of the effective potential \eqref{V3}. To find these non-BPS 
critical points, we set
\begin{equation}
X_{I}:=q_{I}-ZC_{IJK}t^{J}t^{K}  \label{X}
\end{equation}%
Using this notation, we will rewrite \eqref{5deom} in a suitable form \cite{Marrani:2022jpt}. 
To this end, substitute $q_{I}=X_{I}+ZC_{IJK}t^{J}t^{K}$ in \eqref{5deom} to obtain
\begin{equation}
8ZX_{K}+C_{KLM}\big(C^{IL}C^{JM}-C^{IJ}t^{L}t^{M}\big)X_{I}X_{J}=0 \ .
\label{eomxi}
\end{equation}%
Clearly, $X_I=0$ solves this equation trivially and gives BPS critical points. Solutions 
of \eqref{eomxi} with $X_I\neq 0$ correspond to non-BPS critical points. 

To further analyse the above equation, we note that the constraint \eqref{veq1} can be 
rewritten in terms of $X_I$ as 
\begin{equation}
t^{I}X_{I}=0 \ .  \label{tX=0}
\end{equation}%
We can first solve the above constraint to express $X_I$ in terms of $t^I$ and substitute
the resulting values in \eqref{eomxi} which can then be solved to obtain the critical 
points. Care must be taken while solving \eqref{tX=0} for $X_I$, since it remains 
unchanged upon multiplying the $X_I$s by an overall factor. The overall multiplicative 
factor in the $X_I$s can be determined as follows. Multiply both sides of \eqref{eomxi} 
with $C^{KN}X_{N}$ and using \eqref{tX=0} to obtain
\begin{equation}
8ZC^{IJ}X_{I}X_{J}+C_{KLM}C^{KI}C^{LJ}C^{MN}X_{I}X_{J}X_{N}=0\ .
\label{norm}
\end{equation}%
This equation is not homogeneous in $X_I$ and hence will fix the overall multiplicative 
factor in it uniquely. Using the above formalism, models with $h_{1,1}=2$ have been 
extensively analysed in \cite{Marrani:2022jpt}. In the following we will consider 
some explicit examples of Calabi-Yau compactifications with $h_{1,1}=3$ and analyse
the resulting black hole solutions. 

Before we turn our discussion on three parameter Calabi-Yau models, we will outline 
the K\"ahler cone condition, which states that the volume of any effective curve must
be positive:
\begin{equation}
\int_C J > 0 \ , \label{kcnd}
\end{equation}
where $J = t^IJ_I$ is the K\"ahler form and $C$ is any arbitrary effective curve in 
the Calabi-Yau manifold ${\cal M}$. Here $\{J_I\}$ form an integral basis of the cohomology class $H_{1,1}({\cal M},\mathbb{Z})$. Thus, as emphasised in  \cite{Long:2021lon}, we need 
to make sure that the resulting solutions for the attractors must lie within the 
K\"ahler cone.

\section {Three Parameter Model} \label{sec:Three Parameter Model}

Our goal in the present work is to analyze black brane attractors in three parameter 
Calabi-Yau models. In this section we will set up some notations in three parameter
models that will be used throughout this paper to analyze the equations of motion. 
First, for convenience introduce the variables $x,y,z$ to represent the 
K\"ahler moduli $t^1,t^2,t^3$ respectively. Further, introduce the parameters $a,b,\cdots,$
to denote the intersection numbers $C_{IJK}$ such that $C_{111}=a , C_{112}=b, C_{122}=c, 
C_{222}=d, C_{113}=e, C_{123}=f, C_{133}=g, C_{223}=h, C_{233}=i, C_{333}=j$. In addition,
we introduce the functions $A_1,\ldots, A_6$ as 
\begin{eqnarray}\label{a1toa6}
&&A_1 = ax+by+ez\;,\;\;\;\;\;\;\;\;\;  A_2 = bx+cy+fz\;,\;\;\;\;\;\;\;\;\;  A_3 = ex+fy+gz \ , \nonumber \\
&&A_4 = cx+dy+hz\;,\;\;\;\;\;\;\;\;\; A_5 = fx+hy+iz \;,\;\;\;\;\;\;\;\;\;  A_6 = gx+iy+jz \ .
\end{eqnarray}
 With
these notations, the matrix $C_{IJ}=C_{IJK}t^K$ reads as 
\begin{equation}\label{cijmat}
C_{IJ}=%
\begin{pmatrix} A_1 & A_2 & A_3 \cr A_2 & A_4 & A_5 \cr A_3 & A_5 & A_6
\end{pmatrix}%
\ .
\end{equation}%
To express the inverse of the matrix $C_{IJ}$ in an organized way we introduce the functions
$B_1,\ldots,B_6$ as 
 and \begin{eqnarray}\label{b1tob6}
&& B_1 = A_4A_6-A_5^2\;,\;\;\;\;\;\;\;\;\; B_2= A_3A_5-A_2A_6\;,\;\;\;\;\;\;\;\;\;    B_3= A_2A_5-A_3A_4 \ ,   \nonumber \\
&& B_4 = A_1A_6-A_3^2\;,\;\;\;\;\;\;\;\;\;      B_5 = A_2A_3-A_1A_5\;,\;\;\;\;\;\;\;\;\;   B_6 = A_1A_4-A_2^2 \ .
\end{eqnarray}
In terms of these quantities, the matrix inverse $C^{IJ}$ is expressed as 
\begin{equation}\label{cijinv}
C^{IJ}=\frac{1}{A_1B_1+A_2B_2+A_3B_3}%
\begin{pmatrix}
B_1 & B_2 & B_3\cr B_2 & B_4 & B_5\cr B_3 & B_5 & B_6%
\end{pmatrix}%
\ .
\end{equation}%

We further need the expressions for quantities such as $C_I\equiv C_{IJK}t^Jt^K$ and $C^{IJ}q_J$. It 
is straightforward to obtain
\begin{equation}
C_I = C_{IJK}t^{J}t^K=%
\begin{pmatrix}
A_1x+A_2y+A_3z\cr A_2x+A_4y+A_5z\cr A_3x+A_5y+A_6z%
\end{pmatrix} \ , 
\label{cijtj}
\end{equation}%
and
\begin{equation}
C^{IJ}q_J=\frac{1}{A_1B_1+A_2B_2+A_3B_3}%
\begin{pmatrix}
B_1q_1+B_2q_2+B_3q_3\cr B_2q_1+B_4q_2+B_5q_3\cr B_3q_1+B_5q_2+B_6q_3%
\end{pmatrix} \ .
\end{equation}%
The central charge $Z=q_It^I$ is given as 
\begin{equation}
Z = (q_1x+q_2y+q_3z) \ , 
\end{equation}
and the black hole effective potential \eqref{V3} becomes
\begin{equation}\label{veff3p}
V = \frac{3}{2}(q_1x+q_2y+q_3z)^2
  - \frac{(B_1q_1^2+2B_2q_1q_2+2B_3q_1q_3+2B_5q_2q_3+B_4q_2^2+B_6q_3^2)}{2(A_1B_1+A_2B_2+A_3B_3)} \ . 
\end{equation}

\section{BPS Black Holes} \label{BPS Black holes}

We will first analyse the BPS black holes in three parameter Calabi-Yau models. Thus, we 
need to solve the equation \eqref{bpseq} subjected to the constraint \eqref{veq1}. For 
the three parameter models, \eqref{bpseq} becomes 
\begin{eqnarray} \label{bps3p}
 Z (A_1x+A_2y+A_3z) = q_1\ , \cr  Z (A_2x+A_4y+A_5z) = q_2\ , \cr  Z (A_3x+A_5y+A_6z) = q_3\ .
\end{eqnarray}
On the other hand the constraint \eqref{veq1} becomes
\begin{equation} \label{vol3p}
 x (A_1x+A_2y+A_3z) + y (A_2x+A_4y+A_5z) + z (A_3x+A_5y+A_6z) = 1 \ .
\end{equation}
To further analyze the above conditions, note that the left hand sides in equations 
\eqref{bps3p} and \eqref{vol3p} are all degree three homogeneous polynomials in $x,y,z$.
We will introduce the following inhomogeneous coordinates $\tau=x/z$ and $t=y/z$. Further
we define the charge ratios $\rho = q_1/q_3$ and $\sigma = q_2/q_3$. In terms of 
these variables, equations \eqref{bps3p} becomes 
\begin{eqnarray}
z^3 (\rho  \tau +\sigma  t+1) \left(a \tau ^2+2 b \tau  t+c t^2+2 e \tau +2 f t+g\right) = \rho \ , \label{bps1} \\
z^3 (\rho  \tau +\sigma
 t+1) \left(b \tau ^2+2 c \tau  t+d t^2+2 f \tau +2 h t+i\right) = \sigma \ , \label{bps2} \\
z^3 (\rho  \tau +\sigma  t+1) \left(e \tau ^2+2 f \tau  t+2 g \tau +h t^2+2 i
t+j\right) =  1  \ . \label{bps3}
\end{eqnarray}
The constraint on the volume becomes 
\begin{equation}
z^3 (a \tau ^3+3 b \tau ^2 t+3 c \tau  t^2+d t^3+3 e \tau ^2+6 f \tau  t+3 g \tau +3 h t^2+3 i t+j) = 1
\end{equation}
Using the above equation, we can eliminate $z$ in equations \eqref{bps1}-\eqref{bps3}. This
results in three equations in two variables. However, as we will see below, only two of them
remain independent. We need to solve two of these equations for $\tau$ and $t$ in terms of 
the charge ratios $\rho$ and $\sigma$. However, notice that these equations provide two 
coupled cubics in two variables. Thus, it is not possible to obtain an exact analytic 
solution for arbitrary values of the intersection numbers. In order to obtain some insight 
into the problem we will first study the inverse problem,{\it i.e.}, we will solve these
equations for the charge rations $\rho$ and $\sigma$ in terms of the scalar moduli $\tau$
and $t$. Since these equations are linear in $\rho$ and $\sigma$, it is straightforward to
obtain the solution. Consider the first to equations \eqref{bps1} and \eqref{bps2}. We find
\begin{eqnarray}
\rho = \frac{a \tau ^2+2 b \tau  t+c t^2+2 e \tau +2 f t+g}{e \tau ^2+2 f \tau  t+2 g \tau +h t^2+2 i t+j}\ , \cr
\sigma = \frac{b
\tau ^2+2 c \tau  t+d t^2+2 f \tau +2 h t+i}{e \tau ^2+2 f \tau  t+2 g \tau +h t^2+2 i t+j} \ . \label{cr1}
\end{eqnarray}
It can easily be verified that \eqref{bps3} holds true upon substitution of the above 
values for $\rho$ and $\sigma$ in it.

In this paper we will consider some specific examples of three parameter Calabi-Yau manifolds. 
We will consider Calabi-Yau manifolds which are obtained as hypersurfaces from toric 
varieties. A toric variety is given in terms of a reflexive polytope with a specific 
triangulation of its faces. A polytope is reflexive provides it is integral ({\it i.e.}, 
the coordinates of its vertices are all integers) and the corresponding dual polytope is 
also integral. The faces of a reflexive polytope are in one to one correspondence with the 
vertices of the dual polytope. Consider a reflexive polytope with $n$ faces and let the $n$ 
vectors $\vec v_i, (i=1,\cdots, n)$ represent the vertices of the dual polytope. In general, 
they will not all be linearly independent, and, for a polytope in $d$ dimensions, we will 
have $(n-d)$ relations like
\begin{equation}
\sum_{i=1}^n q^r_i \vec{v}_i = 0 \ . \label{wmatrix}
\end{equation}
Here the index $r$ takes values $r=1,\cdots,n-d$. We associate each of the vertices with 
a homogeneous coordinate $z_i\in\mathbb{C}^n$. If we view the coefficients $q^r_i$ in
\eqref{wmatrix} as elements of a $(n-d)\times n$ weight matrix, then each of its row 
defines an equivalence relation among these $n$ homogeneous coordinates with the 
corresponding coefficients as the weights. Thus there are $(n-d)$ such equivalence 
relations altogether. Removing the fixed points and taking quotient with these equivalence 
relations we obtain a $d$ dimensional toric variety. A hypersurface of vanishing first 
Chern class in this toric variety gives a $(d-1)$ dimensional Calabi-Yau manifold.
 
We are interested in Calabi-Yau threefolds. Thus we need to construct toric varieties from
four dimensional reflexive polytopes. All such reflexive polytopes in four dimensions have
been classified by Kreuzer and Sakrke\cite{Kreuzer:2000xy}. The cohomology data for the corresponding toric 
Calabi-Yau manifolds have been computed in \cite{Altman:2014bfa,Altman:2017vzk,Altman:2021pyc}
and the results are listed in the database \cite{CY-database}. In the present work, we will 
use the cohomology data such as the intersection numbers and K\"ahler cone conditions from 
\cite{CY-database} to analyze the black brane configurations.

We will now consider an explicit example of a toric hypersurface Calabi-Yau manifold (THCY).
The polytope ID associated with this THCY as per the Kreuzer-Sakrke classification 
\cite{Kreuzer:2000xy} is 260. In the following we will recapitulate some of the essential cohomology 
data  associated with it from the Calabi-Yau database \cite{CY-database}. The Hodge numbers 
for the Calabi-Yau manifold $\mathcal{M}$ are $h_{1,1}=3$ and $h_{2,1}= 123$ and the 
corresponding Euler number is $-260$. The resolved weight matrix of $\mathcal{M}$ has the 
form
\begin{eqnarray}
Q = \left(
\begin{array}{ccccccc}
 0 & 0 & 0 & 1 & 1 & 3 & 1 \\
 0 & 1 & 1 & 2 & 2 & 6 & 0 \\
 1 & 0 & 0 & 0 & 1 & 2 & 0 \\
\end{array}
\right) \ .
\end{eqnarray}
The volume of the manifold is given by
\begin{equation}
\mathcal{V} =  2 x y^2+2 x y z+4 y^3-4 y z^2 \ .
\end{equation}
From this we can obtain the intersection numbers of the Calabi-Yau manifold  $\mathcal{M}$. We find the 
non-vanishing intersection numbers to be $c=2/3, d=4,f=1/3,i=-4/3$.

Each of the homogeneous coordinates $z_i$ associated with the toric Calabi-Yau $\mathcal{M}$ define 
a holomorphic divisor $D_i$. Denoting the basis elements of $(1,1)$ homology class of $\mathcal{M}$ 
as $\{J_1,J_2,J_3\}$, the divisor classes $D_i$ in terms of them are given as 
$D_1 = - 2 J_1 + J_2 - J_3,
D_2 = J_1, D_3 = J_1, D_4 = 2 J_1 + J_3, D_5 = J_2,
D_6 = 2 J_1 + 2 J_2 + J_3, D_7 = J_3$.
The Mori cone matrix associated with $\mathcal{M}$ is 
\begin{eqnarray}
\mathcal{M}^i_{~j} = \int_{C^i} D_j = C^i\cdot D_j = \left(
\begin{array}{ccccccc}
 0 & 1 & 1 & 0 & 0 & 0 & -2 \\
 1 & 0 & 0 & 0 & 1 & 2 & 0 \\
 -1 & 0 & 0 & 1 & 0 & 1 & 1 \\
\end{array}
\right) \ . 
\end{eqnarray}
The elements of the Mori cone matrix indicate the intersection of generating curves 
$C^i\ (i=1,\ldots,3)$ with the 
toric divisor classes $D_i\ (i=1,\ldots,7)$ \cite{Altman:2017vzk}.
In order to ensure the validity of the K\"ahler cone condition \eqref{kcnd} for our 
attractors, we also need the K\"ahler cone matrix $K$ corresponding to the Calabi-Yau 
manifold. This is the matrix whose elements ${K^i}_j$ are defined by \cite{Altman:2017vzk}
\begin{equation}
{K^i}_j = \int_{C^i} J_j \ . 
\end{equation}
From the expression for the Mori cone matrix we can find the K\"ahler cone matrix. In the 
present case, the K\"ahler cone matrix is given by
\begin{eqnarray}
K = \left(
\begin{array}{ccc}
 1 & 0 & -2 \\
 0 & 1 & 0 \\
 0 & 0 & 1 \\
\end{array}
\right) \ . \label{kkm2}
\end{eqnarray}
This indicates that the K\"ahler moduli $t^I$ for this model will lie inside the K\"ahler cone
provided $x - 2 z > 0, y> 0, z>0$. In terms of the inhomogeneous coordinates these conditions 
become $\tau-2>0, t>0$ and $z>0$. Now, consider the constraint on the volume \eqref{veq1} in
terms of the inhomogeneous coordinates:
\begin{equation} \label{zm1}
2 z^3 t(t+1) (\tau + 2t-2) = 1 \ .
\end{equation}
From the above constraint we observe that $z$ becomes positive in the region $t>0,\tau>2$. Thus,
we do not need to impose it as an additional condition for the solution to remain inside the 
K\"ahler cone.

We will now consider the BPS solutions for this model. Substituting the values of the intersection
numbers in \eqref{cr1} we can obtain the charge ratios in terms of the inhomogeneous coordinates. 
However, in this case the BPS equations $q_I - Z C_{IJK}t^Jt^K=0$ are simple enough to solve directly.
Substituting the values of the intersection numbers and rescaling the charges and variables
appropriately, the equations take the form
\begin{eqnarray}
3 \rho -2 t (t+1) z^3 (\rho  \tau +\sigma  t+1) &=& 0 \ , \cr
3 \sigma -2 z^3 \left(6 t^2+2 \tau  t+\tau -2\right) (\rho  \tau +\sigma  t+1) &=& 0 \ . 
\end{eqnarray}
The variable $z$ in the above equations can be eliminated using the constraint \eqref{zm1}
to obtain
\begin{eqnarray}
\frac{2 \rho  (3 t+\tau -3)-\sigma  t-1}{\tau + 2 t -2} = 0 \ , \
\sigma -\frac{\left(6 t^2+2 \tau  t+\tau -2\right) 
(\rho  \tau +\sigma  t+1)}{3 t (t+1) (\tau + 2 t -2)} = 0 \ .
\end{eqnarray}
These equations can easily be solved to obtain a unique solution for $\tau$ and $t$ 
in terms of the charge ratios $\rho$ and $\sigma$:
\begin{eqnarray} \label{m1sb}
\tau = \frac{24 \rho ^2-4 \rho  (\sigma -3)-\sigma +1}{\rho  (6 \rho -\sigma +2)} \ , \
t = -\frac{2 \rho +1}{6 \rho -\sigma +2} \ .
\end{eqnarray}
The entropy associated with the attractor is 
\begin{equation}\label{entr}
S = \frac{2\pi}{3\sqrt{3}}|Z|^{3/2}
\end{equation}
Upon substituting the solution \eqref{m1sb} in the above and simplifying we find 
\begin{equation}
S = 2^{1/3}\sqrt{3}\pi q_3^2 \Big(\rho(2\rho+1)(4\rho -\sigma+1)\Big)^{2/3} \ .
\end{equation}

We need to make sure that the solution \eqref{m1sb} lies within the K\"ahler cone, {\it i.e.} we 
must have $\tau>2$ and $t>0$. Expressed in terms of the charge ratios, these conditions become
\begin{equation}
\frac{(2\rho + 1)(6 \rho -\sigma +1)}{\rho (6 \rho -\sigma +2)} > 0 \ 
{\rm and} \ \frac{2 \rho +1}{6 \rho -\sigma +2} < 0 \ . 
\end{equation}
To simplify the above conditions further, note that for the attractor solution \eqref{m1sb}, 
we have $\rho^2(\tau + 2 t - 2) = \rho (2\rho+1) (t+1)$. Since both $(\tau + 2 t - 2)$ and 
$(t+1)$ are positive inside the K\"ahler cone, we must have $\rho (2\rho+1)>0$. Thus, we can
have $\rho>0$ or $\rho<-1/2$. Now, requiring $\tau>2$ gives the condition 
\begin{equation}
\frac{6 \rho -\sigma +1}{6 \rho -\sigma +2} > 0 \ .
\end{equation}
Thus, for a given $\rho$ we can either have $6 \rho -\sigma +1>0$ or $6 \rho -\sigma +2<0$. In 
other words, for $\tau>2$ we must have $\sigma>(6\rho+2)$ or $\sigma<(6\rho+1)$. We will now
consider the implication of these two bounds on $\sigma$ on the $t>0$ condition. Note that for 
$\sigma>(6\rho+2)$ we have $(6\rho+2-\sigma)<0$ and hence $t>0$ implies $2\rho+1>0$ and hence
$\rho$ must be positive. For $\sigma<(6\rho+1)$ we have $(6\rho+1-\sigma)>0$ and hence 
$(6\rho+2-\sigma)$ is positive. Thus $2\rho+1<0$ and hence $\rho<-1/2$. To summarize, the solution
\eqref{m1sb} lies within the K\"ahler cone provided $\rho>0, \sigma>(6\rho+2)$ or $\rho<-1/2, \sigma<(6\rho+1)$.

Before we turn our attention to the non-BPS black holes, we will consider one more example of a three parameter THCY model.
The polytope ID associated with this THCY is 230. It has Hodge numbers $h_{1,1}=3, h_{2,1}=111$
and Euler number $\chi=-216$. The resolved weight matrix associated 
with it is given by
\begin{eqnarray}
Q = \left(
\begin{array}{ccccccc}
 0 & 0 & 0 & 0 & 0 & 1 & 1 \\
 0 & 1 & 1 & 1 & 1 & 3 & 0 \\
 1 & 1 & 1 & 0 & 0 & 3 & 0 \\
\end{array}
\right) \ . 
\end{eqnarray}
The volume of the Calabi-Yau manifold $\mathcal{M}$ is given by
\begin{equation}
{\mathcal V} = 2 x^3+2 x^2 y+x^2 z-3 x z^2+9 z^3
\end{equation}
From the above we can read the respective intersection numbers. We find $a=2, b=2/3,
e=1/3,g=-1,j=9$ and all other intersection numbers are zero.
The Mori cone matrix associated with the manifold $\mathcal{M}$ is  
\begin{equation}
\mathcal{M}^i_{~j} = \int_{C^i} D_j  =  \left(
\begin{array}{ccccccc}
 -1 & 0 & 0 & 1 & 1 & 0 & 0 \\
 0 & 0 & 0 & 0 & 0 & 1 & 1 \\
 1 & 1 & 1 & 0 & 0 & 0 & -3 \\
\end{array}
\right) \ , \label{mcm}
\end{equation}
where $\{C^i\}$ are the Mori cone generators. 
 The divisor classes $D_j$
are given in terms of the basis $\{J_1,J_2,J_3\}$ of $(1,1)$ homology class as $D_1 = J_1-J_2,
D_2=J_1,D_3=J_1,D_4=J_2,D_5=J_2,D_6=3J_1-J_3,D_7=J_3$. 
From \eqref{mcm} we find the K\"ahler cone matrix $K$ for $\mathcal{M}$ to be
\begin{equation}
    K=\left(
\begin{array}{ccc}
 0 & 1 & 0 \\
 0 & 0 & 1 \\
 1 & 0 & -3 \\
\end{array}
\right) \ . \label{kkm1}
\end{equation}

We can now study the BPS solutions for this model. Substituting the values of the intersection
numbers in \eqref{cr1} it can be seen that the charge ratios in this case
take the form 
\begin{equation}
\sigma=\frac{2 \tau ^2}{\tau ^2-6 \tau +27}\ , \;\;\;\;\;\;  
\rho=\frac{6 \tau ^2+(4 t+2) \tau -3}{\tau ^2-6 \tau +27} \ .\label{crm1}
\end{equation}
Now we have much a simpler expression and it can be inverted to find the inhomogeneous 
moduli $\tau$ and $t$ in terms of $\sigma$ and $\rho$. We find   
\begin{equation}
\tau=\frac{3}{\sigma-2}\Big(\sigma \pm  \sqrt{2( 3 \sigma - \sigma^2)}\Big)
 \ , \;\;\;\;\;\;
 t=\frac{\rho \tau ^2-6 \rho \tau +27 \rho-6 \tau ^2-2 \tau +3}{4 \tau } \ , \label{seq1}
\end{equation}
for $\sigma\neq 2$. For $\sigma=2$, we have 
\begin{equation}
\tau = \frac{9}{2} \ , \;\;\;\; t = \frac{1}{24}\big(27\rho - 170\big) \ . \label{seq2}
\end{equation}
Upon substituting the solution in the expression for the entropy \eqref{entr} and simplifying, we find 
\begin{eqnarray}
S = \frac{\pi}{12}\left(\frac{2 q_3^3}{3\tau^3}\right)^{1/2}
\frac{\left(\rho ^2 \big(\tau ^2-6 \tau +27\big)- \rho \big(6 \tau ^2+2 \tau -3\big)
+4\tau \big(\sigma  \tau + 1\big) \right)^{3/2}}{\left(\rho\tau\big(\tau^2-6\tau+27\big)-\big(2\tau^3+3\tau-18\big)\right)^{1/2}} \ . 
\end{eqnarray}
Here for simplicity we have expressed the $\sigma$ dependence through $\tau$ as given in \eqref{seq1}.

Several comments are in order. First, since $\tau$ and $t$ are real valued we must have
$3 \sigma - \sigma^2>0$, {\it i.e.}, $\sigma$ lies in the range $0\leq\sigma\leq 3$. 
Further, the values must lie within the K\"ahler cone. From the K\"ahler cone matrix 
\eqref{kkm1}, we find that $y>0, z>0$ and $x - 3 z>0$. Thus, we have $\tau>3$ and $t>0$.
We also need to ensure that $z$ remains positive for both the solutions. We can do so
by examining the volume constraint \eqref{veq1}, which in our model takes the form:
\begin{equation}
z^3 \left(2 \tau ^3+(2 t+1) \tau ^2-3 \tau +9\right) = 1 \ . 
\end{equation}
Notice that the quantity $\left(2 \tau ^3+(2 t+1) \tau ^2-3 \tau +9\right)$ in the left hand
side above remains positive for $\tau>3$ and $t>0$. Thus, imposing the volume constraint 
\eqref{veq1}, $z$ becomes automatically positive provided $\tau>3$ and $t>0$. We do not 
need to impose this condition separately for the solutions to lie inside the K\"ahler cone.

For $\sigma=2$, from \eqref{seq2} we find a unique solution lying within the K\"ahler 
cone provided $\rho > 170/27$. For $\sigma\neq 2$ we have two possible solutions as 
given in \eqref{seq1}. First consider the following solution for $\tau$:
\begin{equation} \label{m1s1}
\tau=\frac{3}{\sigma-2}\Big(\sigma +  \sqrt{2 \sigma ( 3  - \sigma)}\Big)
 \ .
\end{equation}
It may be observed that, for $0\leq\sigma<2$ the value of $\tau$ becomes negative and 
hence lies outside the K\"ahler cone. On the other hand, for $2<\sigma\leq 3$ the value
of $\tau$ is always greater than $3$ and hence lies within the K\"ahler cone. We can
substitute this value of $\tau$ in the expression for $t$ in \eqref{crm1}. We find 
that, for $2<\sigma\leq 3$ the value of $t$ remains positive provided 
\begin{equation}
\rho > \rho_-\equiv \frac{1}{54}\Big((169\sigma - 6) - 4 \sqrt{2 \sigma(3-\sigma)}\Big) \ . 
\end{equation}
For $\sigma=2$, $\rho_-$ takes the minimum possible value $\rho_-=6$. Thus, this branch 
of BPS attractors exists for $2<\sigma\leq 3$ and $\rho>6$.

We will now consider the second solution for $\tau$ in \eqref{seq1}:
\begin{equation} \label{m1s2}
\tau=\frac{3}{\sigma-2}\Big(\sigma - \sqrt{2 \sigma ( 3  - \sigma)}\Big) 
 \ .
\end{equation}
It is easy to notice that, for $1<\sigma\leq 3$ the value of $\tau$ remains greater 
than $3$ and it lies within the K\"ahler cone. Substituting the above expression for 
$\tau$ in the expression for $t$ in \eqref{seq1} we find that it remains positive 
provided
\begin{equation}
\rho > \rho_+ \equiv \frac{1}{54}\Big((169\sigma - 6) + 4 \sqrt{2 \sigma(3-\sigma)}\Big) \ . 
\end{equation}
Since $\rho_+$ monotonically increases with $\sigma$, it takes the minimum possible value
$\rho_+=19/6$ at $\sigma=1$. This branch exists for $1<\sigma\leq 3$ and $\rho>19/6$.

From the above we can see that, while no solution exists for $\sigma<1$, we have a unique
solution described by \eqref{m1s2} and the corresponding $t$ when $\rho > \rho_+$ and 
$\sigma$ takes values in the range $1<\sigma\leq 2$. On the other hand, for $2<\sigma<3$
there is a possibility of existence of both the solutions. Notice that $\rho_+\geq\rho_-$, 
with the equality holding for $\sigma=3$. Thus, we have a unique solution described by
\eqref{m1s1} (and the corresponding solution for $t$) if $2<\sigma\leq 3$ and 
$\rho_-<\rho<\rho_+$, whereas both the solutions exist if $\rho>\rho_+$.

As a concrete example, take $\sigma=3/2$ and $\rho=5$. In this case the equations of 
motion gives rise to a unique solution inside the K\"ahler cone
\begin{equation}
\tau = 9 \big(\sqrt{2}-1\big) \ , \;\;\; t = \frac{1}{12}\big(19\sqrt{2}-23\big) \ , \;\; {\rm and} \;\;
z = \left(\frac{2}{9\big(1887 \sqrt{2} -2641\big)}\right)^{1/3} \ . 
\end{equation}
Similarly, consider the values $\sigma=5/2$ and $\rho=38/5$. We find 
\begin{equation}
\tau = 3 \big(5 + \sqrt{10}\big) \ , \;\;\; t= \frac{1}{300}\big(13\sqrt{10} - 35\big) \ ,
\;\;\; z = \left(\frac{30267}{2} + \frac{46773}{\sqrt{10}}\right)^{-1/3} \ .
\end{equation}
On the other hand, if we take $\sigma=5/2$ and $\rho=8$ we obtain two solutions both lying
inside the K\"ahler cone:
\begin{eqnarray}
&& \tau = 3 \big(5 + \sqrt{10}\big) \ , \;\; t= \frac{1}{60} (65 + 17 \sqrt{10}) 
\ , \;\; z = \left(\frac{2}{9(3627 + 1121 \sqrt{10})}\right)^{1/3} \ , \cr
{\rm and} && \tau = 3 \big(5 - \sqrt{10}\big) \ , \;\; t= \frac{1}{60} (65 - 17 \sqrt{10})
\ , \;\; z = \left(\frac{2}{9(3627 - 1121 \sqrt{10})}\right)^{1/3}. 
\end{eqnarray}

One distinguished feature is the existence of two distinct solutions for a given value
of $\sigma$ and $\rho$ in the range $2<\sigma\leq 3$ and $\rho>\rho_+$.
It is rather interesting to find multiple solutions preserving supersymmetry for the same 
set of charges in our model. Such examples were constructed first in the case of five 
dimensional supergravity theories arising from the compactification of M-theory on two 
parameter Calabi-Yau manifolds. It was later realized that the multiple solutions lie in 
different disconnected branches of the moduli space. While the examples studied in 
\cite{Marrani:2022jpt} admitted multiple non-BPS attractors, the BPS solutions for a 
given set of charges were all unique. Though multiple non-BPS attractors are found to
occur in many supergravity theories, the existence of multiple BPS attractors seems 
quite exceptional.

\section{Non BPS Black Holes}\label{sec:Non BPS Black Holes}

We will now turn our attention to non-BPS black holes in the three parameter Calabi-Yau
models. Thus, we need to analyze the equation of motion
\begin{equation}
8ZX_{K}+C_{KLM}\big(C^{IL}C^{JM}-C^{IJ}t^{L}t^{M}\big)X_{I}X_{J}=0 \ ,
\label{eomxi1}
\end{equation}%
with $X_I\neq 0$. We need to keep in mind that $X_I$'s satisfy  the constraint \eqref{tX=0},
$t^IX_I=0$. Setting $X_I = \hat{X} \tilde{X}_I$ in \eqref{norm}, we obtain
\begin{equation}
\hat{X}=-\frac{8ZC^{IJ}\tilde{X}_{I}\tilde{X}_{J}}{C_{KLM}C^{KP}C^{LQ}C^{MN}%
\tilde{X}_{P}\tilde{X}_{Q}\tilde{X}_{N}} \ .
\end{equation}%
Using the above in the definition of $X_I$ from \eqref{X}, we find 
\begin{equation}
q_I = Z \left( C_{IJ}t^J 
-\frac{8 \tilde{X}_I C^{JK}\tilde{X}_{J}\tilde{X}_{K}}{C_{KLM}C^{KP}C^{LQ}C^{MN}%
\tilde{X}_{P}\tilde{X}_{Q}\tilde{X}_{N}}  \right) \ . \label{qieq}
\end{equation}
In terms of the charge ratios $\rho=q_1/q_3$ and $\sigma=q_2/q_3$ the above equation
can be rewritten as 
\begin{eqnarray}\label{nbpsrho}
\rho&=&\frac{C_{1J}t^J C_{KLM}C^{KP}C^{LQ}C^{MN} \tilde{X}_{P}\tilde{X}_{Q}\tilde{X}_{N}
- 8 \tilde{X}_1 C^{JK}\tilde{X}_{J}\tilde{X}_{K}}
{C_{3J}t^J C_{KLM}C^{KP}C^{LQ}C^{MN} \tilde{X}_{P}\tilde{X}_{Q}\tilde{X}_{N}
- 8 \tilde{X}_3 C^{JK}\tilde{X}_{J}\tilde{X}_{K}} \ , \\
\sigma&=&\frac{C_{2J}t^J C_{KLM}C^{KP}C^{LQ}C^{MN} \tilde{X}_{P}\tilde{X}_{Q}\tilde{X}_{N}
- 8 \tilde{X}_2 C^{JK}\tilde{X}_{J}\tilde{X}_{K}}
{C_{3J}t^J C_{KLM}C^{KP}C^{LQ}C^{MN} \tilde{X}_{P}\tilde{X}_{Q}\tilde{X}_{N}
- 8 \tilde{X}_3 C^{JK}\tilde{X}_{J}\tilde{X}_{K}} \ . \label{nbpsgma}
\end{eqnarray}

With $\tilde{X}_I$ being algebraic functions of $t^I$ the right hand sides of the above 
equations are only functions of the moduli. A consistent solution to the above conditions 
leads a non-BPS black hole. To find whether the resulting configuration is stable, we need 
to consider the recombination factor \cite{Long:2021lon}. The recombination factor $R$ is 
defined as the ratio of the black hole mass to the mass of the $M2$-brane wrapping the 
minimum piece-wise calibrated curve in the same homology class as the non-holomorphic curve 
corresponding to the black hole:
\begin{equation}
R = \frac{M_C}{M_{C^\cup}} \ ,
\end{equation}
where $M_C$ is the mass of the $M2$ brane wrapping the non-holomorphic curve $C$ and 
$M_{C^\cup}$ is the mass of the $M2$ brane wrapping the minimum volume piece-wise 
calibrated curve $C^\cup$ in the homology class $[C]$. The mass of the non-BPS black
hole is given by $M_C = \sqrt{V_{\rm cr}}$, the square root of the black hole 
potential at the critical point. On the other hand, if $C = \sum\alpha_IC^I$, then
$M_{C^\cup} = \sum|\alpha_I|t^I$. Thus,
\begin{equation}
R = \frac{\sqrt{V_{\rm cr}}}{\sum|\alpha_I|t^I} \ . \label{recomb}
\end{equation}
If this quantity is greater than one, the black hole becomes unstable and decays into 
the constituent BPS/non-BPS pairs. Whereas, for $R<1$ recombination of the brane/anti-brane
pairs takes place and the non-BPS black hole becomes stable \cite{Long:2021lon}.

Let us now focus on the constraint \eqref{tX=0}. In the case of two parameter Calabi-Yau
models \cite{Marrani:2022jpt} it took the form $\tilde{X}_1 x + \tilde{X}_2 y = 0.$
Thus, for the two parameter case it was possible to obtain the expressions 
for $X_I$ uniquely upon  solving this constraint. However, this is not the case for models with $h_{1,1}>2$. In the 
present case the constraint takes the form $\Tilde{X}_1 x +\Tilde{X}_2 y +\Tilde{X}_3 z 
=0$. In terms of the rescaled coordinates we have $\Tilde{X}_1\tau +\Tilde{X}_2 t 
+ \tilde{X}_3=0$, and hence we can at most express $\tilde{X}_3$ in terms of $t^I$ and
$\tilde{X}_1,\tilde{X}_2$. Thus, to obtain these quantities we need to solve the 
full equations of motion. In the following we will consider examples of three parameter models
where we will explicitly solve the equations of motion to obtain the non-BPS attractors.

\subsection{Examples}\label{BHcase1}

We will now consider non-BPS attractors for the both the examples considered in
the previous section. Let us first consider the first example, say the model 1.
We need the explicit expression for the black hole effective potential to obtain
the non-BPS solutions. Substituting the intersection numbers for the model 1 in
\eqref{veff3p} we find
\begin{equation} \label{veffm2}
V = \frac{3}{4 y (y+z) (x+2 y-2 z)} \big( v_{11} q_1^2 + v_{22} q_2^2 + v_{33} q_3^2
+ 2 v_{12} q_1 q_2 + 2 v_{23} q_2 q_3 + 2 v_{31} q_3 q_1\big) \ , 
\end{equation}
where the functions $v_{ij}$ are defined as follows
\begin{eqnarray}
v_{11} &=& 2 x^3 y (y+z)+x^2 \left(4 y^3-4 y z^2+1\right)+8 x (y-z)+16 \left(3 y^2+z^2\right), \cr
v_{22} &=& y^2 \left(2 x y (y+z)+4 y^3-4 y z^2+1\right), \cr
v_{33} &=& 2 y^2 \left(x z^2+2\right)+y \left(2 x z^3-4 z^4+4 z\right)+4 y^3 z^2+z^2, \cr
v_{12} &=& y \left(2 x^2 y (y+z)+x \left(4 y^3-4 y z^2-1\right)-8 y\right), \cr
v_{23} &=& y \left(2 x y^2 z+y \left(2 x z^2-4 z^3-2\right)+4 y^3 z-z\right), \cr
v_{31} &=& 2 y^2 \left(x^2 z+6\right)+2 y z \left(x^2 z-2 x z^2+4\right)+4 x y^3 z+z (4 z-x) \ .
\end{eqnarray}
The constraint on the volume \eqref{veq1} for the Calabi-Yau manifold $\mathcal{M}$ becomes
\begin{equation}\label{vconm2}
2 x y^2+2 x y z+4 y^3-4 y z^2 = 1 \ .
\end{equation}

In order to obtain the critical points, we need to extremize the black hole effective potential \eqref{veffm2} using the method of 
Lagrange multiplier to incorporate the constraint \eqref{vconm2} on the scalar fields.
Upon eliminating the Lagrange multiplier we obtain two independent equations which 
after some simplifications take the form
\begin{eqnarray}
q_1^2 f_{11} + q_2^2 f_{22} + 2 q_3^2 y(y+z)(2y+z) + f_{12} q_1 q_2 
+ f_{23} q_2 q_3 + f_{31} q_3 q_1 = 0 \ , \cr
 \Big(q_1 \left(2 x^2 y (y+z)+x \left(4 y^3-4 y z^2+1\right)
 +8 y\right) + q_2 y \left(2 x y
(y+z)+4 y^3-4 y z^2-1\right) \cr 
+q_3 \left(2 x y^2 z+2 y \left(x z^2-2 z^3+1\right)+4 y^3 z+z\right)\Big)  (q_1 (x-4 z) - q_3 (y+z)) = 0 \ .
\end{eqnarray}
The functions $f_{ij}$ are defined as 
\begin{eqnarray}
f_{11} &=& 2 x^3 y \left(2 y^2+3 y z+z^2\right)+x^2 \left(20 y^4+16 y^3 z-12 y^2 z^2+y \left(2-8 z^3\right)+z\right) \cr &+& 2 x \left(12
y^5-16 y^3 z^2+5 y^2+4 y z \left(z^3-1\right)-3 z^2\right)-8 \left(3 y^3+9 y^2 z+y z^2-z^3\right),\cr
f_{22} &=& y^2 (y+z)\text{  }\left(2 x y (y+z)+4 y^3-4 y z^2+1\right), \cr
f_{12} &=& y \left(2 x^2 y^2 (y+z)+x y \left(16 y^3+12 y^2 z-8 y z^2-4 z^3-1\right) \right. \cr
&+& \left. 2 \left(12 y^5-16 y^3 z^2+5 y^2+4 y z \left(z^3+2\right)+z^2\right)\right), \cr
f_{23} &=& y (y+z)\left(2 x y^2 z+2 x y z^2+4 y^3 z-4 y \left(z^3+1\right)-z\right), \cr
f_{31} &=& 4 y^3 \left(x^2 z+4 x z^2-8 z^3-6\right)-2 y^2 z \left(-3 x^2 z+6 x z^2+19\right)
\cr &+& 2 y z \left(x^2 z^2-4 x z^3-x+4 z^4-4 z\right)+20
x y^4 z+z^2 (2 z-x)+24 y^5 z .
\end{eqnarray}
We can rewrite these equations in terms of the inhomogeneous coordinates and the charge 
ratios and eliminate $z$ using the constraint \eqref{vconm2}. The equations motion simplify 
remarkably to have the form
\begin{eqnarray}
(\rho  \tau +4 \rho  t+t+1)(\rho  (\tau -4)-t-1) =0, \cr
t^3 g_1 + t^2 (g_2 - 8\tau\rho^2) + t (4 \rho ^2+4 \rho +1) - 2 \tau t \rho^2 (\tau-2)-\rho ^2 (\tau -2)^2 =0,
\end{eqnarray}
where $g_1$ and $g_2$  are functions of the charge ratios $\rho$ and $\sigma$ with the 
expression
\begin{eqnarray}
g_1 = 12 \rho ^2+\rho  (12-8 \sigma )+\sigma ^2-2 \sigma +2 \ , \  
g_2 = 36 \rho ^2 -8 \rho  (\sigma -2)+\sigma ^2-2 \sigma +3 \ .
\end{eqnarray}

It is indeed possible to solve these equations exactly to obtain the inhomogeneous 
coordinates in terms of the charge ratios. They admit four independent solutions. They
are given as 
\begin{eqnarray}
&& \tau =-\frac{8 \rho ^2+8 \rho -\sigma +1}{\rho  (2 \rho -\sigma )},t=\frac{2 \rho +1}{2 \rho -\sigma } , \label{nbp1} \\
&& \tau =\frac{8 \rho ^2+\sigma -1}{\rho  (6 \rho -\sigma +2)},t=-\frac{2 \rho +1}{6 \rho -\sigma +2} , \label{nbp2} \\
&& \tau =\frac{8 \rho ^2-4 \rho  \sigma +4 \rho -\sigma +1}{2 \rho ^2-\rho  \sigma },t=\frac{2 \rho +1}{2 \rho -\sigma } , \label{nbp3} \\
&& \tau =\frac{24 \rho ^2-4 \rho  (\sigma -3)-\sigma +1}{\rho  (6 \rho -\sigma +2)},t=-\frac{2 \rho +1}{6 \rho -\sigma +2} \ . \label{nbp4}
\end{eqnarray}
The last of these solutions corresponds to BPS black holes. They have been analyzed in detail
in \S\ref{BPS Black holes}. On the other hand, the first three solutions do not satisfy the
BPS equations. They corresponds to non-BPS attractors. 

We need to make sure that the non-BPS solutions lie within the K\"ahler cone. From the 
expression of the K\"ahler cone matrix we observe that the values of $t$ and $\tau-2$ must 
be positive. Let us analyze this condition for the first solution given in \eqref{nbp1}.
We find 
\begin{equation}\label{nbp1kk}
\frac{2 \rho +1}{2 \rho -\sigma } > 0 \ {\rm and} \ 
\frac{(2\rho+1)(6\rho - \sigma+1)}{\rho (2\rho - \sigma)}<0 \ .
\end{equation}
Further, we consider the combination $(t+1)(\tau+2t-2)$ which is positive inside the 
K\"ahler cone. For \eqref{nbp1}, this implies $\rho(2\rho+1)<0$. Thus, the value of $\rho$
must lie in the range $-1/2<\rho<0$. Since $2\rho+1$ is positive, the first inequality
in \eqref{nbp1kk} implies $2\rho-\sigma>0$ and hence $\sigma<2\rho$. The second inequality
in \eqref{nbp1kk} now implies $\sigma<6\rho+1$. Thus, the solution \eqref{nbp1} remains
inside the K\"ahler cone provided $-1/2<\rho<0$ and $\sigma< {\rm min}\{2\rho,6\rho+1\}$.

We will now focus on the K\"ahler cone condition for the solution \eqref{nbp2}. A similar
analysis tells that once again the values of $\rho$ lies in the range $-1/2<\rho<0$. 
Requiring $t>0$ for a given $\rho$ in this range now gives the bound $\sigma>6\rho+2$.
Requiring $\tau-2>0$ gives $\sigma>2\rho+1$. Thus, the solution \eqref{nbp2} lies inside
the K\"ahler cone provided $-1/2<\rho<0$ and $\sigma> {\rm max}\{2\rho+1,6\rho+2\}$. 
Finally, let us consider the solution \eqref{nbp3}. In this case $\rho$ takes values 
in the range $\rho>0$ or $\rho<-1/2$. For $\rho>0$ we must have $\sigma<2\rho$ and
for $\rho<-1/2$ the value of $\sigma$ lies in the range $\sigma>2\rho+1$. For these 
values of $\rho$ and $\sigma$ the solution \eqref{nbp3} remain inside the K\"ahler 
cone. Further, these bounds ensure that all the four branches of solutions as given in
\eqref{nbp1}-\eqref{nbp4} are mutually exclusive from each other. Thus, in this model 
both the BPS as well as non-BPS attractors are unique.

Let us now turn our attention to the second model consider in \S\ref{BPS Black holes}.  
The black hole effective potential \eqref{veff3p} for this model takes the form
\begin{eqnarray}
V = \frac{3 q_1}{2 x} \left(q_1 x^3+2 x^2 (q_2  y + q_3 z) - q_2 \right)  
 + \frac{1 }{8 x^2 (x-9 z)} \left( q_2^2 V_3 + q_2 q_3 V_2 + q_3^2 V_1\right) \ .
 \label{vnbpsm1}
\end{eqnarray}
where, for easy reading we have introduced the functions
\begin{eqnarray}
V_1 &=& 4 x (3 x z^2-27 z^3+1) \ , \cr
V_2 &=& 4 x  \left(6 x^2 y z-54 x y z^2-x+3 z\right) \ ,  \cr
V_3 &=& 12 x^3 y^2 - x^2 \left(108 y^2 z - 19\right)+3 x (2 y-55 z)-18 z (3 y+z) \ . 
\end{eqnarray}
The constraint on the volume \eqref{veq1} becomes 
\begin{equation}
2 x^3+x^2 (2 y+z)-3 x z^2+9 z^3 = 1 \ . \label{veq1m1}
\end{equation}
The equations of motion is obtained upon extremizing the  potential \eqref{vnbpsm1} subject to 
the constraint \eqref{veq1m1}. We find

\begin{eqnarray}
24 q_1^2 x^5 (x-9 z)^2- q_2  f_3(x,y,z) \big(4 x^2 (q_1 x+q_2 y +q_3 z)+q_2\big)
+4 q_2 q_3 x^2 f_1(x,y,z) \ {~} && \cr 
+12 q_1 x^2 (x-9 z)^2 (2 q_2 x^2 y+q_2+2 q_3 x^2 z)-q_2^2 f_5(x,y,z) 
-4 q_3^2 x^4  =  0 \ ,  &&  \\
(q_2 f_1(x,y,z)-2 q_3 x^2)
\big(2 x^2 (2 q_1 x (x-9 z)^2+q_3 f_2(x,y,z))+q_2 f_4(x,y,z)\big) =  0 \ . && \ .
\end{eqnarray}
In the above the functions $f_1,\cdots,f_5$ are defined as follows
\begin{eqnarray}
f_1(x,y,z) &=& x^2-6 x z+27 z^2 \ , \cr
f_2(x,y,z) &=& 2 x^2 z-36 x z^2+162 z^3+3 \ , \cr
f_3(x,y,z)&=& 3 (x-9 z)^2(6 x^2+2 x (2 y+z)-3 z^2)\ , \cr
f_4(x,y,z) &=& 4 x^4 y-72 x^3 y z+x^2 (324 y z^2-1)-18 x z+81 z^2 \ , \cr
f_5(x,y,z) &=& x \big(19 x^3+6 x^2 (2 y-55 z)-27 x z (8 y-53 z)+324 z^2 (3 y+z)\big) \ . 
\end{eqnarray}
We express these equations in terms of the rescaled variables $\tau=x/z, t=y/z$ and 
eliminate $z$ using \eqref{veq1m1} to obtain 
\begin{eqnarray}
&& 4 \tau ^4 \left(2 \tau ^3+(2 t+1) \tau ^2-3 \tau +9\right)
- 12 \rho  (\tau -9)^2 \tau^2 \left(2\tau^2 (1 + \rho\tau)
- \sigma \left(4 \tau ^3+\tau ^2-9\right)\right) \cr
&& - 4 \sigma\tau ^2 \left(2 \tau ^5+(2 t-29) \tau^4+(363-24 t) \tau^3 
+9 (30 t-143) \tau ^2-27 (36 t+29) \tau +972\right)
 \cr && 
+ \sigma^2\left(74 \tau ^7+\left(96 t^2-3408 t+4785\right) \tau ^5 
-9 \left(192 t^2-1588 t-777\right) \tau^4 
+54 \left(144 t^2+148 t\right.\right. \cr && \left.\left.
-271\right) \tau^3 -(1259-194 t) \tau ^6 
-81 (174 t-263) \tau ^2+2187 (8 t+5) \tau -6561\right) = 0 \ ,
\end{eqnarray}
and 
\begin{eqnarray}
&& \left(2 \tau ^2-\sigma  \left(\tau ^2-6 \tau +27\right)\right)
\Big(2 \tau ^2 \left(6 \tau ^3+(6 t+5) \tau ^2-45 \tau +189\right)+4
\rho  (\tau -9)^2 \tau ^3  \cr && 
- \sigma\left(2 \tau ^5-(2 t-37) \tau ^4+3 (36 t-49) \tau ^3-18 (27 t+7) \tau ^2+405 \tau -729\right)\Big) = 0 .
\end{eqnarray}

Even though these equations look complicated, it is indeed to solve the inverse problem 
exactly to express the charge ratios in terms of the inhomogeneous coordinates $(\tau,t)$.
We find three independent solutions:
\begin{eqnarray} \label{m1bps1}
\rho &=& \frac{6 \tau ^2+(4 t+2) \tau -3}{\tau ^2-6 \tau +27} \ ,
 \;\;\;\;\;\;  \sigma = \frac{2 \tau ^2}{\tau ^2-6 \tau +27}\ , \\
 \rho &=& -\frac{2 \tau ^3+2 (2 t+1) \tau ^2-9 \tau +36}
 {\tau  \left(\tau ^2-6 \tau +27\right)}
 \ , \;\;\;\;\;\; \sigma = \frac{2 \tau ^2}{\tau ^2-6 \tau +27}\  \label{m1nbps1}
\end{eqnarray}
and
\begin{eqnarray} \label{m1nbps2}
\rho&=&-\frac{328 \tau ^5+36 \left(6 t^2+4 t-63\right) \tau ^3+(548 t+310) \tau ^4+(9981-1134 t) \tau ^2+81 (84 t+23) \tau +729}{92 \tau ^5+3 (30 t-763) \tau ^4-81 (28 t-121) \tau ^3+162 (63 t+19) \tau ^2-3645 \tau +6561} \cr
\sigma&=&-\frac{2 \tau ^2 \left(52 \tau ^3+27 (2 t+3) \tau ^2-567 \tau +1701\right)}{92 \tau ^5+3 (30 t-763) \tau ^4-81 (28 t-121) \tau ^3+162 (63 t+19)\tau ^2-3645 \tau +6561} 
\end{eqnarray}

The first solution \eqref{m1bps1} corresponds to BPS black holes which has been 
considered in \S\ref{BPS Black holes}. On the other hand solutions obtained in 
\eqref{m1nbps1} and \eqref{m1nbps2} give rise to non-BPS black holes. We will now 
analyze them in detail. Consider first \eqref{m1nbps1}. We can solve them to obtain
analytic expressions for the inhomogeneous coordinates $(\tau,t)$. Denoting 
\begin{eqnarray} \label{ttau}
\tau_\pm &=& \frac{3 \left(\sigma \pm \sqrt{2 \sigma (3-\sigma ) }\right)}{(\sigma -2)} \ , \cr
t_\pm &=& \frac{1}{36 (\sigma -2)^3 \sigma }\Big(12 (405 \rho -11) \sigma ^2-8 (2187 \rho +31) \sigma -59 \sigma ^4+222 \sigma ^3+192\cr
&\mp & \sqrt{2\sigma(3-\sigma )} \left(972 \rho  \sigma +5832 \rho +55 \sigma ^3-222 \sigma ^2+228 \sigma -8\right)\Big) \ ,
\end{eqnarray}
there are two independent solutions $(\tau_+,t_+)$ and $(\tau_-,t_-)$. The expression for 
$\tau$ in these solutions remain the same as that of BPS black holes as given in \eqref{seq1}.
Thus, from the discussion followed by \eqref{m1s2} we understand that requiring $\tau_-$ to lie inside the K\"ahler
cone for the solution $(\tau_-,t_-)$ gives rise to the constraint $1<\sigma\leq 3$ for the 
possible values of $\sigma$. Further, introducing $\rho_\pm$ as 
\begin{equation*}
\rho_\pm=\frac{1}{26244\sigma}
\Big(\sigma (185 \sigma^2-948 \sigma-396) 
\pm 2 \sqrt{2 \sigma (3-\sigma ) }(167 \sigma^2+18 \sigma+72)\Big) \ ,
\end{equation*}
we find that the value $t_-$ in the solution $(\tau_-,t_-)$ remains positive if $\rho<\rho_-$.
At $\sigma=3$, $\rho_-$ takes its maximum possible value $\rho_-^m= -175/2916$. Thus, the
non-BPS configurations $(\tau_-,t_-)$ exist for $1<\sigma\leq 3$ and $\rho<-175/2916$.
Let us now focus on the solution described by $(\tau_+,t_+)$. In this case, for 
$2<\sigma\leq 3$ the value of $\tau_+$ remains greater than $3$. Requiring $t_+>0$ in
this solution, we find that $\rho$ must satisfy the upper bound $\rho<\rho_+$. From the 
expression for $\rho_+$ we find that $\rho_+\leq 0$ throughout with the equality holding
at $\sigma=2$. Thus, this branch exists when $2<\sigma\leq 3$ and $\rho<0$. Further, 
we notice that  $\rho_-\leq\rho_+$ in the range $2<\sigma\leq 3$. Thus, for $2<\sigma\leq 3$
and $\rho_-<\rho<\rho_+$ only $(\tau_+,t_+)$ lies within the K\"ahler cone whereas, for
$\rho<\rho_-$ both the solutions survive.

Before we turn our attention to \eqref{m1nbps2},
we would like to compare the BPS and non-BPS solutions we have obtained so far. Though
both the set of solutions exist only when $1<\sigma\leq 3$, the range of $\rho$ is entirely
different. While, for the BPS solutions the minimum possible value of $\rho$ is $19/6$,
for the non-BPS solutions $\rho$ must take negative values. Thus the BPS solutions obtained
from \eqref{m1bps1} and non-BPS solutions derived from \eqref{m1nbps1} do not coexist
simultaneously.

We will now consider \eqref{m1nbps2}. This involves coupled equations of degree five 
in $(\tau,t)$ and hence it is not possible to obtain an exact analytical expression
for these variables in terms of $(\rho,\sigma)$. Nevertheless, it is possible to 
obtain the range of charge ratios which admit attractor solutions lying inside the 
K\"ahler cone. First thing to notice is that there exist curves of singularities in
the moduli space which do not correspond to any possible extremal black hole 
configuration for finite values of black hole charges. These curves are obtained by
setting the denominators in the expressions for $\rho$ and $\sigma$ in \eqref{m1nbps2}
to zero. We find 
\begin{equation}
92 \tau ^5+3 (30 t-763) \tau ^4-81 (28 t-121) \tau ^3+162 (63 t+19) \tau ^2-3645 \tau +6561 = 0 \ .
\end{equation}
This equation can be rewritten as 
\begin{equation}
t = -\frac{92 \tau ^5-2289 \tau ^4+9801 \tau ^3+3078 \tau ^2-3645 \tau +6561}{18 \tau ^2 \left(5 \tau ^2-126 \tau +567\right)} \ . \label{tttau}
\end{equation}
The above expression has two poles as well as two zeros. Denote the poles as $\tau_{p1},
\tau_{p2}$ and the zeros as $\tau_{z1}, \tau_{z2}$. The values of $\tau_{p1}$ and $\tau_{p2}$
are 
$$\tau_{p1} = \frac{9}{5}\big(7-\sqrt{14}\big)\simeq 5.865 \ , \;\;\;
\tau_{p1} = \frac{9}{5}\big(7+\sqrt{14}\big)\simeq 19.335 \ , $$
and the zeros $\tau_{z1},\tau_{z2}$ are approximately 
$\tau_{z1} \simeq 5.859, \ \tau_{z2} \simeq 19.266$. The value of $t$ as given in
\eqref{tttau} remains positive in the intervals $(\tau_{z1},\tau_{p1})$ and $(\tau_{z2},
\tau_{p2})$. Thus, there are two curves of
singularities. They form a pair of almost straight lines which start respectively 
from $\tau_{z1}$ and $\tau_{z2}$ at $t=0$ and asymptote to $\tau=\tau_{p1}$ and 
$\tau=\tau_{p2}$ lines as $z\rightarrow\infty$. Any point inside the K\"ahler cone
which does not lie on either of these two curves correspond to an extremal non-BPS 
black hole configuration for some suitably chosen charges.

The singularity curves naturally divide the moduli space into three regions. We denote 
the region bounded by the singularity curve asymptote to the line $\tau=\tau_{p1}$ and 
the $\tau=3$ boundary of the K\"ahler cone as region 1, the region bounded by the curve 
asymptote to the line $\tau=\tau_{p2}$ and the boundary $\tau=\infty$ as region 2 and 
the region bounded by the two curves of singularities as region 3. If we treat $\sigma$
and $\rho$ as functions of $(t,\tau)$, then they do not possess any extremum in regions
1 and 2. In region 1, $\sigma$ takes a maximum value $-79/257$ at the boundary point
$t=0,\tau=3$. As we move away from this point, $\sigma$ becomes more and more negative
and $\sigma\rightarrow -\infty$ as we approach the singularity curve. In region 2,
$\sigma$ attains a maximum value of $-26/23$ in the limit $\tau\rightarrow\infty$. It 
takes a smaller value for any point interior to this region and as we approach the
singularity curve, $\sigma\rightarrow -\infty$. In region 3, $\sigma\rightarrow\infty$ 
as we approach either of the singularity curves from an interior point of this region.
However, it remains positive throughout inside the region 3 and hence form a valley
shaped surface. To find the minimum value of $\sigma$ in this region we set both the
partial derivatives $\frac{\partial\sigma}{\partial t}$ and $\frac{\partial\sigma}
{\partial\tau}$ to zero to obtain:
$$ (\tau - 9)^5 = 0 $$
and 
\begin{eqnarray*}
(\tau -9) \Big(729 \left(28 t^2+36 t+143\right) \tau ^4+8 (6 t+2635) \tau ^6+144 (276 t+17) \tau ^5 \cr -1458 (30 t+217) \tau ^3+6561
(4 t+13) \tau ^2-275562 \tau +413343\Big) = 0 \ .
\end{eqnarray*}
Thus $\sigma$ has a line of minima given by $\tau=9$, with the minimum value $\sigma_m=3$.

We now focus on the range of values for $\rho$ that admits attractor solutions. In region 1
as well as region 2 it does not admit an extremum. At the boundary point $t=0,\tau=3$ of 
region 1 it takes the maximum value $\rho = -575/514$. As we move away from this point, it
becomes more negative and diverges as we approach the singularity curve. Similarly, in 
region 2, it has a maximum value $\rho=-82/23$ in the limit $\tau\rightarrow\infty$. It 
becomes more negative at finite $\tau$ and as we approach the singularity curve, $\rho$
diverges. In the interior of region 3, $\rho$ takes positive values and $\rho\rightarrow
\infty$ as we approach the singularity curves from any point in the interior. To find 
the minimum value of $\rho$ in this region, we set $\frac{\partial\rho}{\partial t}$ 
and $\frac{\partial\rho}{\partial\tau}$ to zero. A numerical computation indicates that
these equations do not admit any solution in the interior of region 3. For any fixed $t$
it admits a minimum for some $\tau$. The value of this minimum decreases with $t$. Thus,
the functional form of $\rho$ defines a valley which slopes downward towards smaller 
values of $t$. To find the smallest allowed value of $\rho$, we set $t=0$ in it and
extremize. We find 
\begin{eqnarray}
&& \Big(1640 \tau ^4+1240 \tau ^3-6804 \tau ^2+19962 \tau +1863\Big) 
\Big(92 \tau ^5-2289 \tau ^4+9801 \tau ^3+3078 \tau ^2 \cr &&
-3645 \tau +6561\Big)- \Big(460 \tau ^4-9156 \tau ^3+29403 \tau ^2+6156 \tau -3645\Big) \Big(328 \tau ^5+310 \tau ^4 \cr && 
-2268 \tau ^3+9981 \tau ^2+1863 \tau +729\Big) = 0 \ .
\end{eqnarray}
This admits a unique solution $\tau_m$ in region 3. Numerically we find $\tau_m\simeq
9.03$. The value of $\rho$ corresponding to this point is $\rho_m\simeq 9.277$.

From the above analysis we find that, for $\sigma>3$ and $\rho>\rho_m$ we have multiple
non-BPS solutions in region 3. If $-26/23<\sigma<-79/257$ and $-82/23<\rho<-575/514$ we 
have a unique non-BPS attractor in region 1 for a given set of charges. Finally, for 
$\sigma<-26/23$ and $\rho<-82/23$ we have two non-BPS attractors for a given set of 
charges one of which lies in region 1 and the other in region 2. No attractors exist
for charges lying in the range $-79/257<\sigma<3$ and $-575/514<\rho<\rho_m$. Further, 
we note that since the above solutions exist only when $\sigma>3$ or $\sigma<-79/257$, 
whereas the solutions obtained from \eqref{m1bps1} and \eqref{m1nbps1} exist only when 
$1<\sigma\leq 3$, all the three set of solutions are mutually exclusive.

As an example, consider the values $\sigma=4, \rho=16$. It gives a pair of solutions for
$(\tau,t,z)$ with approximate values $(7.101,2.231,10.225)$ and $(12.281,4.102,17.682)$, 
both lying in region 3. If we take $\sigma=-2,\rho=-7$ we find one solution in region 1 
with $\tau\simeq 4.853$ and $t\simeq 0.61$ and another in region 2 with $\tau\simeq 55.008$
and $t\simeq 10.363$. In this case $z$ takes the approximate values $6.506$ and $73.587$ 
respectively. On the other hand, for $\sigma=-1,\rho=-7/2$ we find a unique 
solution in region 1 with $\tau\simeq 4.298, t\simeq 0.383$ and $z\simeq 5.724$.

\subsection{Stability}\label{stbl1}

Now we will analyze the stability of the non-BPS black holes obtained in the previous 
subsection. For this purpose we need to obtain the recombination factors of the respective
black hole configurations. Let the non-BPS black hole results from wrapping a $M2$ brane on  
a non-holomorphic curve $C=\alpha_1 C^1 + \alpha_2 C^2 + \alpha_3 C^3$ of the Calabi-Yau manifold 
$\mathcal{M}$. The charges $q_I$ of the black hole are given by
\begin{equation}
q_I = \int_C J_I  = \alpha_1 C^1\cdot J_I + \alpha_2 C^2\cdot J_I + \alpha_3 C^3\cdot J_I \ . 
\end{equation}
The intersection numbers $C^I\cdot J_L$ can be obtained from the expression for the K\"ahler cone
matrix \eqref{kkm2} corresponding to the THCY manifold $\mathcal{M}$. We find 
\begin{equation}
q_1 = \alpha_1 - 2 \alpha_3 \ , \ q_2 = \alpha_2 \ , \ q_3 = \alpha_3 \ .
\end{equation}
From the above we find the coefficients $\alpha_I$ in terms of the black hole charges $q_J$
as $\alpha_1 = q_1 + 2 q_3 = (2\rho+1)q_3, \alpha_2 = q_2=\sigma q_3$ and $\alpha_3 = q_3$. 
Thus the recombination factor \eqref{recomb} corresponding to the non-BPS black holes in 
model 1 have the expression
\begin{equation}
R = \frac{\sqrt{V_{\rm cr}}}{z|q_3| \left(|(2\rho+1)|\tau + |\sigma| t + 1\right)} \ .
\end{equation}
Here $V_{\rm cr}$ is the value of the effective black hole potential at the critical point. 
Substituting the expression for the effective black hole potential \eqref{veffm2}, using
the volume constraint \eqref{vconm2} and simplifying, we find
\begin{equation} \label{recom1}
R = \frac{\sqrt{3\left(\rho ^2 \left(\tau ^2-4 \tau +8\right)+4 \rho +t^2 \left(24 \rho ^2+\rho  (12-8 \sigma )+\sigma ^2-2 \sigma +2\right)+t
\left(4 \rho ^2 \tau +8 \rho +2\right)+1\right)}}{\left(|(2\rho+1)|\tau + |\sigma| t + 1\right)}
\end{equation}

We need to substitute the solutions in the three branches of non-BPS black holes to obtain
the corresponding recombination factors in each of the cases. Consider the solution 
\eqref{nbp1} first. Recall that, the allowed values of $\rho$ and $\sigma$ for this solution
are negative, whereas $2\rho+1>0$. Keeping this in mind and substituting \eqref{nbp1} in
\eqref{recom1} we find the recombination factor to have the form
\begin{equation}
R = -\frac{3 \rho  (2 \rho +1) (4 \rho -\sigma +1)}{16 \rho ^3+2 \rho ^2 (\sigma +11)+10 \rho -\sigma +1} \ .
\end{equation}
The black hole becomes unstable when $R>1$. This happens for 
\begin{equation}\label{comp}
\sigma>\frac{40 \rho ^3+40 \rho ^2+13 \rho +1}{4 \rho ^2+3 \rho +1} \ .
\end{equation}
However, recall that for the solution \eqref{nbp1} to remain inside the 
K\"ahler cone we need to impose the bound $\sigma<{\rm min}\{2\rho,6\rho+1\}$. This may not 
be always compatible with \eqref{comp}. We find that \eqref{comp} does not hold when $\rho$
takes values in the range $(\sqrt{17}-9)/32<\rho<0$ or $-1/2<\rho<(\sqrt{17}-9)/16$. For 
these values of $\rho$ the recombination factor $R$ is always smaller than one and the 
corresponding black hole solutions are stable. For $(\sqrt{17}-9)/16<\rho<(\sqrt{17}-9)/32$
we can choose $\sigma$ to satisfy \eqref{comp}. In that case the resulting non-BPS 
black holes become unstable. However, we can also choose $\sigma$ not to satisfy this 
bound. The non-BPS black holes for those values of $\sigma$ remain stable against decay 
into constituent BPS/anti-BPS pairs.

Next consider the second branch of solutions given by \eqref{nbp2}. Here the charge ratios
take values in the interval $-1/2<\rho<0$ and $\sigma>{\rm max}\{2\rho+1,6\rho+2\}$. Hence 
the values of $\sigma$ and $2\rho+1$ always remain positive. Substituting \eqref{nbp2} in 
\eqref{recom1} and simplifying we find 
\begin{equation}
R = -\frac{3 \rho  (2 \rho +1) (4 \rho -\sigma +1)}{16 \rho ^3-2 \rho ^2 (\sigma -7)+\sigma -1} \ .
\end{equation}
In order to get $R>1$ the charge ratios need to satisfy the bound
\begin{equation}\label{comq}
\sigma>\frac{40 \rho ^3+32 \rho ^2+3 \rho -1}{8 \rho ^2+3 \rho -1} \ .
\end{equation}
However, for $-1/2<\rho<0$, the rhs of the above equation is always smaller than 
${\rm max}\{2\rho+1,6\rho+2\}$. Since the K\"ahler condition implies $\sigma>{\rm max}\{2\rho+1,6\rho+2\}$, all such values of $\sigma$ satisfy the bound \eqref{comq}. Thus, 
the recombination factor is always  greater than one and hence the resulting non-BPS 
black holes in this branch are all unstable.

Finally, for the branch \eqref{nbp3}, the recombination factor takes the form
\begin{equation} \label{frmla}
R = \left|\frac{3 \rho  (2 \rho +1) (4 \rho -\sigma +1)}{| 2 \rho +1|  \left(8 \rho ^2-4 \rho  (\sigma -1)-\sigma +1\right)+\rho  (2 \rho
 | \sigma | +| \sigma | +2 \rho -\sigma )}\right| \ .
\end{equation}
For this solution to be inside the K\"ahler cone we can have either $\rho>0,\sigma<2\rho$ 
or $\rho<-1/2,\sigma>2\rho+1$. We will treat the stability conditions for $\rho>0,
0<\sigma<2\rho$ and $\rho>0,\sigma<0$ separately. Consider first the case $\rho>0,
0<\sigma<2\rho$. Analyzing the formula \eqref{frmla} we find that for $\rho>\left(2-\sqrt{2}\right)^{1/3} \left.\left(2^{1/3}+\left(2+\sqrt{2}\right)^{2/3}\right)\right/4$ the value 
of the recombination factor always becomes greater than one irrespective of the value of 
$\sigma$ and the black holes become unstable. However, for $0<\rho<\left(2-\sqrt{2}\right)^{1/3} \left.\left(2^{1/3}+\left(2+\sqrt{2}\right)^{2/3}\right)\right/4$ the black holes become 
unstable if $\sigma>1-8\rho^3/(3\rho+1)$, and stable if  $\sigma<{\rm min}\{
2\rho,1-8\rho^3/(3\rho+1)\}$. In the case $\rho>0,\sigma<0$, black holes become stable 
if $\sigma<{\rm min}\{0,(1+3\rho-8\rho^3)/(1+5\rho+4\rho^2)\}$ and unstable otherwise.
Similarly, for $\rho<-1/2,\sigma>0$ we find stable non-BPS black holes provided 
$\sigma>{\rm max}\{0,(1+3\rho-4\rho^2-8\rho^3)/(1+3\rho+4\rho^2)\}$. For $\rho<-1/2,
2\rho+1<\sigma<0$ there are no stable attractors if $\rho<\rho_0$ where $\rho_0\simeq -0.776$
is a root of $8\rho^3 + 4 \rho^2 - 3 \rho - 1 =0$. For $\rho_0<\rho<-1/2$ we have stable 
attractors if $\sigma$ takes values in the interval ${\rm max}\{2\rho+1,-(8\rho^3 + 4 \rho^2 - 3 \rho - 1)/(\rho+1)\}$.

We will now analyze the stability of the non-BPS black holes in model 2. Let the corresponding $M2$
brane wraps a non-holomorphic curve $C=\sum \alpha_I C^I$.
Using the expression for the K\"ahler cone matrix in \eqref{kkm1}, we find the charges $q_I$
of the black hole to be
\begin{equation}
q_1 = \alpha_3 \ , \;\; q_2 = \alpha_1 \ , \;\; {\rm and} \;\; q_3 = \alpha_2 - 3\alpha_3  \ . 
\end{equation}
Thus, the coefficients $\alpha_I$ are given in terms of the charges $q_I$ as 
$\alpha_1 = q_2=\sigma q_3, \alpha_2= 3q_1 + q_3 = (3\rho+1)q_3$ and $\alpha_3=q_1=\rho q_3$.
The recombination factor \eqref{recomb} can be expressed as 
\begin{equation}
R = \frac{\sqrt{V_{\rm cr}}}{z|q_3|\big(|\sigma|\tau + |(3\rho+1)| t + |\rho|\big) } \ . 
\end{equation}
Using the constraint \eqref{veq1m1} in \eqref{vnbpsm1}, the recombination factor takes the form
\begin{equation}
R = \frac{1}{2\tau\big(|\sigma|\tau + |(3\rho+1)| t + |\rho|\big)} 
\sqrt{\frac{\widetilde{V}}{2 (\tau-9)}} \ , \label{rmast}
\end{equation}
where we have introduced 
\begin{eqnarray}
\widetilde{V} &=& \sigma ^2 \Big(38 \tau ^5-311 \tau ^4-258 \tau ^3+24 t^2 (\tau -9) \tau ^2+648 \tau ^2 \cr
&+&2 t \left(25 \tau ^4-216 \tau ^3-54 \tau^2+108 \tau -243\right)-1431 \tau -162\Big) \cr
&+&4 \tau ^2 \left(3 \rho ^2 (\tau -9) \tau ^2+6 \rho  (\tau -9) \tau +2 \tau ^3+(2 t+1) \tau ^2-18\right) \cr
&-& 4 \sigma  \tau  \Big(3 \rho  \left(2 \tau ^4-17 \tau ^3-12 \tau ^2+36 \tau -81\right)
+2 \tau ^4+(2 t-5) \tau ^3 \cr
&-&6 (2 t+1) \tau ^2+18 (3 t+1)\tau -27\Big) \ .
\end{eqnarray}
The black hole becomes stable for $R<1$. By directly substituting specific values of charges and 
the corresponding solutions $t$ and $\tau$ we find stable black holes for a wide range of charges.
In the following we consider a class of configurations for stable as well as unstable black holes.

We choose a specific value of the charge ratio $\sigma=3/2$. In this case, the equations 
of motion admit a unique solution given by $(\tau_-,t_-)$ in \eqref{ttau}. Substituting 
$\sigma=3/2$ in $(\tau_-,t_-)$ we find the solution to have the simple form 
\begin{equation}\label{3by2}
\tau_- = 9 \left(\sqrt{2}-1\right) \;\; {\rm and} \;\;
t_- = \frac{1}{36} \left(141-161 \sqrt{2}\right)-3 \left(\sqrt{2}-1\right) \rho .
\end{equation}
For $t_->0$ we need $\rho$ to satisfy the bound $\rho<-(181+20\sqrt{2})/108\simeq -1.938$. We 
substitute these
values for $\tau_-$ and $t_-$ (and $\sigma=3/2)$) in \eqref{rmast} to obtain the following simple 
expression for the recombination factor:
\begin{equation*}
R = \frac{9 \left(1+\sqrt{2}\right) \sqrt{11664 \left(17 - 12 \sqrt{2}\right) \rho ^2+216 \left(1952 \sqrt{2}-2765\right) \rho -317524
\sqrt{2}+449721}}{2 \left(324 \left(\sqrt{2}-1\right) \rho ^2+\left(591 \sqrt{2}-567\right) \rho +647 \sqrt{2}-627\right)} \ . 
\end{equation*}
It can be easily verified that $R<1$ provided 
\begin{equation}
\rho< -\frac{1}{216} \left(367+8 \sqrt{2}+\sqrt{5 \left(28101+80 \sqrt{2}\right)}\right) 
\simeq -3.49 \ . 
\end{equation}
Thus, for $\sigma=3/2$ we obtain non-BPS attractors as given by \eqref{3by2}. The K\"ahler cone 
condition requires that $\rho\lesssim -1.938$. These non-BPS black holes remain unstable in the 
range $-3.49\lesssim\rho\lesssim -1.938$ and become stable when $\rho\lesssim -3.49$.

We will now take the value $\sigma=5/2$. In this case there are two solutions 
$(\tau_+,t_+)$ and $(\tau_-,t_-)$ with 
\begin{equation} \label{stables}
\tau_\pm = 3 \left(5\pm\sqrt{10}\right)  , \ 
t_\pm =  - \frac{1}{180} \left(108 \left(5\pm\sqrt{10}\right) \rho 
\pm 271 \sqrt{10}+1423\right) \ .
\end{equation}
The solution $(\tau_-,t_-)$ exists when $\rho<-(4405 + 68\sqrt{10})/1620$ and 
the solution $(\tau_+,t_+)$ exists when $\rho<-(4405 - 68\sqrt{10})/1620$. Thus, 
we have multiple solution for $\rho<-(4405 + 68\sqrt{10})/1620$. We can substitute
these solutions in \eqref{rmast} to find the respective recombination factors
\begin{equation*}
R_\pm = \frac{15 \sqrt{5} \sqrt{6480 \left(89\pm 28 \sqrt{10}\right) \rho ^2-216 \left(13375 \pm 4208 \sqrt{10}\right) \rho \pm 1138324 \sqrt{10}+3618013}}{2
\left(540 \left(7\pm 2 \sqrt{10}\right) \rho ^2+3 \left(3595\pm 1026 \sqrt{10}\right) \rho \pm 5426 \sqrt{10}+19025\right)}
\end{equation*}
Introducing $\rho_{s\pm}$ as 
\begin{equation*}
\rho_{s\pm} = -\frac{8695\mp 8 \sqrt{10}+\sqrt{87877865\pm 72560 \sqrt{10}}}{3240} \
\end{equation*}
we find that $R_\pm<1$ provided $\rho<\rho_{s\pm}$. Hence, the solution $(\tau_-,t_-)$ is
unstable for  $\rho_{s-}<\rho<-(4405 + 68\sqrt{10})/1620$ and becomes stable for $\rho<\rho_{s-}$
whereas the solution $(\tau_+,t_+)$ is unstable for $\rho_{s+}<\rho<-(4405 - 68\sqrt{10})/1620$
and becomes stable for $\rho<\rho_{s+}$. Since $\rho_{s-}<\rho_{s+}$, both the solutions in \eqref{stables} become stable when $\rho<\rho_{s-}$.

We will now analyze the stability of solutions for \eqref{m1nbps2}. In this case it is not possible
to obtain any analytical results. We can numerically solve the equations of motion for various 
choices of $\rho$ and $\sigma$ and obtain the recombination factor for each of the solutions. As 
we have noted previously, in this case there exist two singularity curves which divide the moduli 
space into three regions. We first consider $\sigma>3$ and $\rho>\rho_m$ for which the solutions 
lie in region 3. We took the value $\rho=45$ and varied $\sigma$ from $4$ to $14$. In each of 
these cases, there are two distinct solutions. Upon computing the recombination factor we observe
that for a fixed $\rho$ and for small values of $\sigma$ the black holes become stable. The 
recombination factor increases monotonically and the black holes become unstable for larger
values of $\sigma$. We also observe that, for fixed values of $(\rho,\sigma)$ with a smaller 
value $\sigma$, the recombination factors corresponding to the two distinct solutions almost
coincide with each other. In the following, we plot the solutions for $\tau$ and $t$ in 
Fig.\ref{figtt}. We can clearly see two branches of solutions in both the cases. Since the 
recombination factors corresponding to the two branches of solutions are close to each other, 
we plot it for the lower branch of solutions. Further, we plot the difference of the two 
recombination factors separately (see Fig.\ref{figrr}).

\begin{center}
\includegraphics[width=0.49\linewidth]{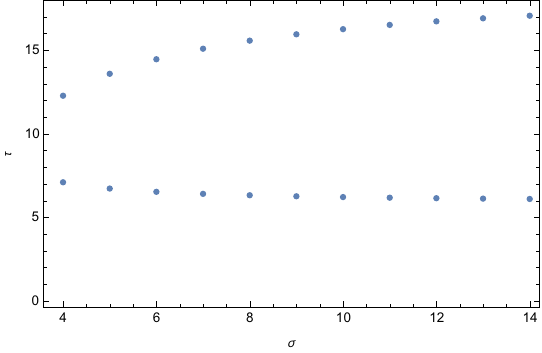}
\includegraphics[width=0.49\linewidth]{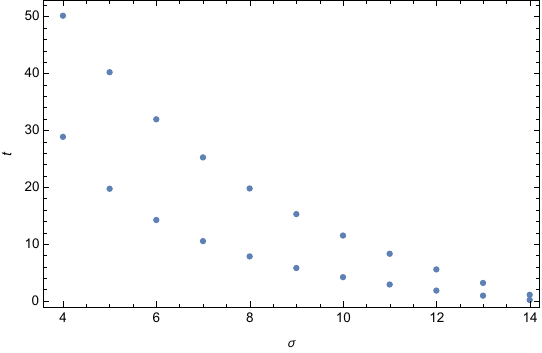}
\captionof{figure}{Two branches of solutions for $\tau$ and $t$ with $\rho=45$
and  $4\leq\sigma\leq 14$.} \label{figtt}
\end{center}

\begin{center}
\includegraphics[width=0.49\linewidth]{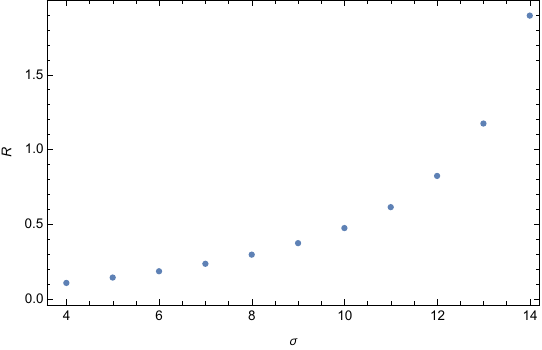}
\includegraphics[width=0.49\linewidth]{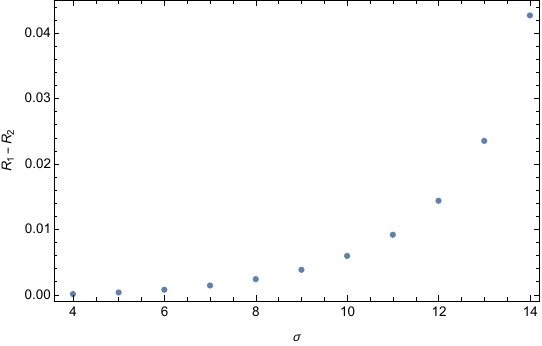}
\captionof{figure}{Recombination factor for the lower branch of solutions and the difference
 of two recombination factors. The recombination factors take slightly higher values in the 
 upper branch of solutions.} \label{figrr}
\end{center}

We will now consider the solutions in regions 1 and 2. In this case, we have a unique non-BPS 
attractor in region 1 when $\sigma$ and $\rho$ lies in the range $-26/23<\sigma<-79/257$ and 
$-82/23<\rho<-575/514$ and multiple non-BPS attractors when $\sigma<-26/23,\rho<-82/23$. We 
will first focus on the unique solutions in region 1. We chose $\sigma$ to be valued in the 
range $-21/20\leq\sigma\leq -11/20$ and $\rho=-7/2$. The resulting solution for $\tau$ and $t$ 
are plotted in Fig.\ref{ttaur1}. We notice that $t$ monotonically increases with $\sigma$, 
while $\tau$ decreases as we increase
it. We have numerically computed the corresponding recombination factors (see Fig.\ref{rrr}). We 
find stable non-BPS black holes for smaller values of $\sigma$.

Further, we choose the value $\rho=-15$ for various $\sigma$ in the range 
$-23/5\leq\sigma\leq -13/5$. Multiple solutions occur for these values of the charge ratios.
We find one solution in region 1 and the other in region 2. The solutions for $(\tau,t)$ 
in region 1 are plotted in Fig.\ref{tauu} and solutions for $(\tau,t)$ in the region 2 are plotted in
Fig.\ref{tttu}. The corresponding recombination factors are plotted in Fig.\ref{rrr}.
The lower curve in Fig.\ref{rrr} corresponds to recombination factors in region 1 and 
upper curve in it corresponds to recombination factors in region 2. The black hole solutions
with a larger value of $\sigma$ remain stable. The value of $R$ increases gradually as we
decrease $\sigma$ with a fixed value of $\rho$ and the black holes become unstable after 
a critical value.

\begin{center}
\includegraphics[width=0.49\linewidth]{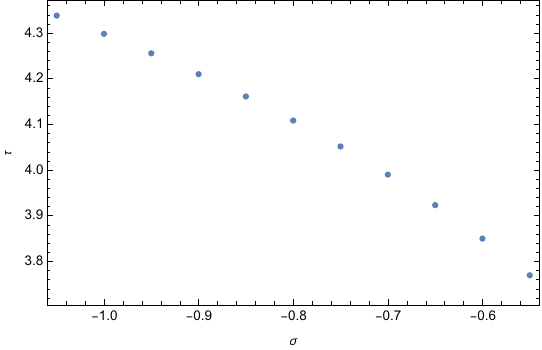}
\includegraphics[width=0.49\linewidth]{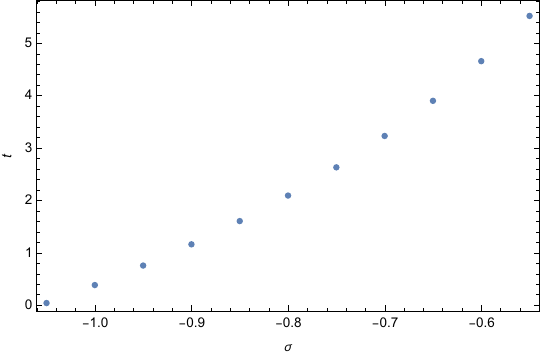}
\captionof{figure}{Solutions for the inhomogeneous coordinates $\tau$ and $t$ in region 1
for $\rho=-7/2$ and $-21/20\leq\sigma\leq -11/20$.} \label{ttaur1}
\end{center}

\begin{center}
\includegraphics[width=0.49\linewidth]{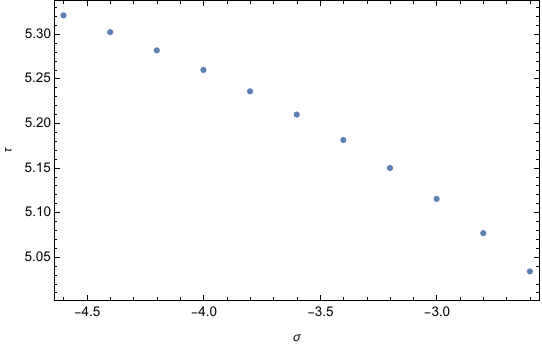}
\includegraphics[width=0.49\linewidth]{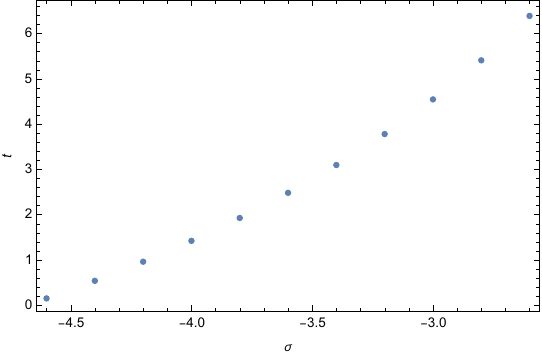}
\captionof{figure}{Solutions for the inhomogeneous coordinates $\tau$ and $t$ in region 1
for $\rho=-15$ and $-23/5\leq\sigma\leq -13/5$.} \label{tauu}
\end{center}
\begin{center}
\includegraphics[width=0.49\linewidth]{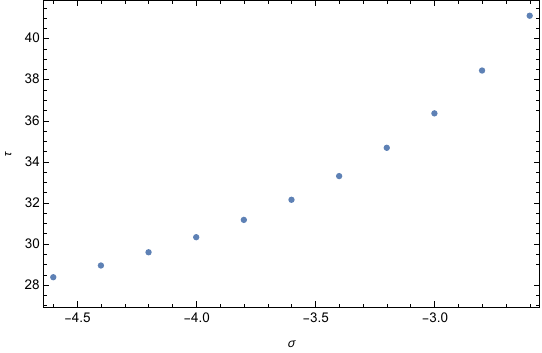}
\includegraphics[width=0.49\linewidth]{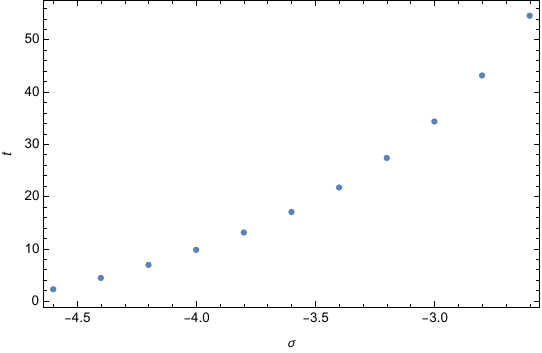}
\captionof{figure}{Solutions for the inhomogeneous coordinates $\tau$ and $t$ in region 2
for $\rho=-15$ and $-23\leq\sigma\leq -13/5$.} \label{tttu}
\end{center}
\begin{center}
\includegraphics[width=0.49\linewidth]{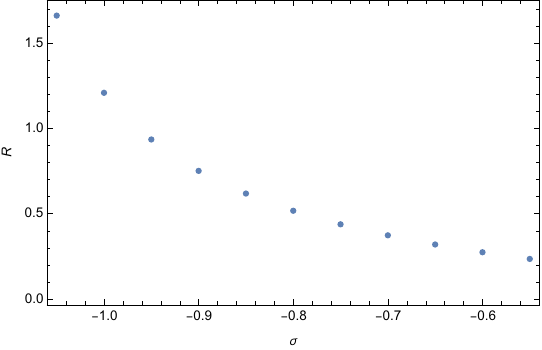}
\includegraphics[width=0.49\linewidth]{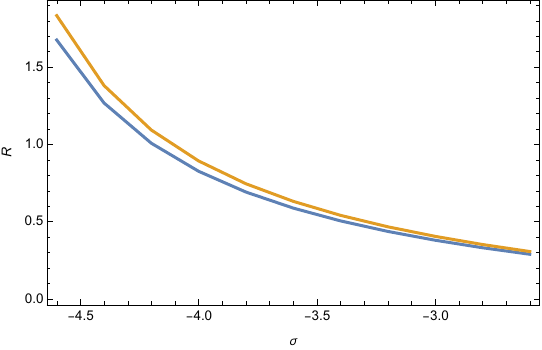}
\captionof{figure}{Recombination factors for the above solutions. The figure in the left
corresponds to recombination factors for solutions in region 1 with $\rho=-7/2$ and $-21/20\leq\sigma\leq -11/20$. The figure in right corresponds to recombination factors for the multiple solutions
with $\rho=-15$ and $-23\leq\sigma\leq -13/5$.  } \label{rrr}
\end{center}

\section{Non-BPS black strings}\label{Non Bps black strings}

We will now consider black strings in our supergravity theory. These are magnetically charged 
objects with charges $p^I$. They are obtained upon wrapping $M5$ branes on a four cycle of 
the Calabi-Yau manifold. The central charge associated with the black string is given by
  \begin{equation}\label{ccrg}
      Z=C_{IJK}p^It^Jt^K \ .
  \end{equation}
The effective potential corresponding to the black string is 
  \begin{equation} \label{effpt}
      V=3Z^2-2C_{IJ}p^Ip^J \ .
  \end{equation}
To obtain the equation of motion, we need to extremize this effective black string potential.
Note that the moduli $t^I$ are constrained to satisfy \eqref{veq1}. We can use the method 
of Lagrange multipliers to extremize the potential. The resulting equation of motion is given
by
\begin{equation}
6Z(p^{J}-Zt^{J})-C^{JK}C_{KLM}p^{L}p^{M}+C_{LM}p^{L}p^{M}t^{J}=0 \ .
\end{equation}%
Upon introducing $X^{I}\equiv p^{I}-Zt^{I}$, the above equation can be rewritten as \cite{Marrani:2022jpt}
\begin{equation}
4ZX^I + X^JX^KC_{JKL}\big(t^Lt^I - C^{LI}\big) = 0 .  \label{nbpsbs}
\end{equation} 
Setting $X^I=0$ we obtain BPS solutions. Upon simplification, they take the form 
\begin{equation}
t^{I}=\frac{p^{I}}{\big(C_{JKL}p^{J}p^{K}p^{L}\big)^{1/3}} \ .
\end{equation}
Thus, for any five dimensional supergravity the resulting BPS black string solutions are 
unique. Solutions of the equation of motion \eqref{nbpsbs} with $X^I\neq 0$ correspond to
non-BPS black strings. 

For non-BPS black strings, using \eqref{veq1} we can show that $X^I$ satisfies the 
constraint
\begin{equation}
C_{IJK}t^{I}t^{J}X^{K}=0 \ .  \label{jjj}
\end{equation}%
Multiplying $X^I$ on \eqref{nbpsbs} we can show that 
\begin{equation}
4ZC_{IJ}X^IX^J - C_{IJK}X^IX^JX^K = 0 .  \label{scal}
\end{equation}
Setting $X^I=\check{X}\tilde{X}^{I}$ in the above we find 
\begin{equation}
\check{X}=\frac{4ZC_{MN}\tilde{X}^{M}\tilde{X}^{N}}{C_{IJK}\tilde{X}^{I}%
\tilde{X}^{J}\tilde{X}^{K}}.
\end{equation}
and $\tilde{X}^I$ satisfies the constraint $C_{IJK}t^{I}t^{J}\tilde X^{K}=0$

In the present work we will focus on three parameter models. Using the expression for 
$C_{IJK}t^Jt^K$ given in \eqref{cijtj} we find the central charge \eqref{ccrg} to have the form
\begin{equation}
Z = \big(A_1x+A_2y+A_3z\big)p^1 + \big(A_2x+A_4y+A_5z\big)p^2+\big(A_3x+A_5y+A_6z\big) p^3 \ .
\end{equation}
Similarly, using \eqref{cijmat} we find
\begin{equation}
C_{IJ}p^Ip^J = A_1(p^1)^2 + A_4 (p^2)^2 + A_6 (p^3)^2 + 2\big(A_2p^1p^2+A_3p^1p^3+A_5p^2p^3\big) \ .
\end{equation}

Substituting these values in \eqref{effpt} we can obtain the expression for the black string
effective potential. We will once again focus on the explicit examples we have considered in
the previous sections to study BPS and non-BPS black holes. Let us consider the model 1 
first. The effective black string potential for this model is given by
\begin{eqnarray}
V &=&\frac{4}{3} \left(p^1 y (y+z)+ p^2 x (2 y+z)+6 p^2 y^2-2 p^2 z^2+p_3 y (x-4 z)\right)^2 \cr 
&-& \frac{4}{3} \left(p^1 p^2 (2 y+z)+p^3 y (p^1-2 p^3)+(p^2)^2 (x+6 y)+p^2p^3 (x-4 z)\right)
\end{eqnarray}
Upon extremization with respect to the scalar fields $x,y,z$ subject to the constraint 
\eqref{vconm2} we find the equations of motion to be of the form
\begin{eqnarray}
&&(y+z) \left(p^2 (p^1-4 p^3)-2 (y (p^1-4 p^3)+p^2 (x-4 z)) 
\left(p^1 y (y+z)+p^2 x (2 y+z)\right.\right. \cr 
&+& \left.\left. 6 p^2 y^2-2 p^2 z^2+p^3 y (x-4 z)\right)\right)
+(x-4 z) \left(2 (p^2 (2 y+z)+p^3 y) \left(p^1 y (y+z) \right.\right. \cr &+& \left.\left. 
p^2 x (2 y+z)
+6 p^2 y^2-2 p^2 z^2+p^3 y (x-4 z)\right)-p^2 (p^2+p^3)\right)=0, \cr
&& y (x-4 z) \left(2 (p^1 (2 y+z)+2 p^2 (x+6 y)+p^3 (x-4 z)) 
\left(p^1 y (y+z)+p^2 x (2 y+z) \right.\right. \cr &+& \left.\left. 6 p^2 y^2-2 p^2
z^2+p^3 y (x-4 z)\right)-2 p^1 p^2-p^3 (p^1-2 p^3)-6 (p^2)^2\right)+
 \left(x (2 y+z) \right. \cr &+& \left. 6 y^2-2 z^2\right)
 \left(p^2 (p^1-4 p^3)-2 (y (p^1-4 p^3)+p^2 (x-4 z)) \left(p^1 y
(y+z) \right.\right. \cr &+& \left.\left. p^2 x (2 y+z) +
6 p^2 y^2-2 p^2 z^2+p^3 y (x-4 z)\right)\right)=0 . \label{eombs}
\end{eqnarray}
As before we will introduce the inhomogeneous coordinates $\tau=x/z, t=y/z$ and eliminate
$z$ using the constraint \eqref{vconm2}. In addition, we will rescale the magnetic charges 
$p^I$ as $p^1 = r p^3$ and $p^2 = s p^3$. The equations of motion \eqref{eombs} simplify a 
lot upon expressing in terms of the inhomogeneous coordinates $\tau,t$ and charge ratios 
$r,s$. We find
\begin{eqnarray}
 (r t+r-s \tau +4 s-4 t-\tau ) (r t+r+4 s t+s \tau +\tau -4)=0, \cr
t (\tau -4) \left((2 r t+r+2 s (6 t+\tau )+\tau -4) 
\left(t (r t+r+\tau -4)+s \left(6 t^2+2 \tau  t+\tau -2\right)\right) \right. \cr
 \left. -t (t+1) \left(2 r
s+r+6 s^2-2\right) (2 t+\tau -2)\right)+ \left(6 t^2+2 \tau  t+\tau -2\right) 
\left((r-4) s t (t+1)\right.  \cr  \left. (2 t+\tau -2)-((r-4) t+s (\tau -4)) \left(t (r t+r+\tau -4)+s \left(6 t^2+2
\tau  t+\tau -2\right)\right)\right)=0 \ . \
\end{eqnarray}
It is possible to solve the above equations exactly to find the attractor values of the 
moduli in terms of the magnetic charges. Apart from the BPS solution $\tau=r, t=s$, there
are three additional solutions to the above equations. These solutions correspond to 
non-BPS black strings. They have the simple form
\begin{eqnarray}
\tau &=& 4-r-4 s \ , \ t =  s \ , \label{br1st1} \\
\tau &=& \frac{r+8 s}{1+2 s} \ , \ t = -\frac{s}{1+2s}\ , \label{br2st1} \\
\tau &=& \frac{4-r+4 s}{1+2 s} \ , \ t = -\frac{s}{1+2 s}\ . \label{br3st1}
\end{eqnarray}
Requiring the solution to satisfy the K\"ahler cone condition $\tau>2,t>0$ we find 
the supersymmetric solutions for $r>2,s>0$. For the first branch of non-BPS solutions
\eqref{br1st1}, we have $s>0,r<-2(2s-1)$. For the second branch \eqref{br2st1} we 
have $-1/2<s<0$ and $r>2(1-2s)$. For the third branch $-1/2<s<0$ and $r<2$. All
these branches of solutions are mutually exclusive from each other. Moreover, the
resulting solution for a given set of charges is always unique.

Now we will analyze non-BPS black strings in model 2. For this case, the effective
black string potential takes the form
\begin{equation}
V = \frac{1}{3}\Big( \tilde{f}_1 (p^1)^2 + 4 x^4 (p^2)^2 + \tilde{f}_2 (p^3)^2
+ \tilde{f}_3 p^1 p^2 + \tilde{f}_4 p^1 p^3 + \tilde{f}_5 p^2p^3 \Big) \ , 
\end{equation}
where we have introduced the functions 
\begin{eqnarray}
\tilde{f}_1 &=& 36 x^4+24 x^3 (2 y+z)+16 x^2 \left(y^2+y z-2 z^2\right)-12 x \left(2 y z^2+z^3+1\right)-4 y+9 z^4-2 z, \cr
\tilde{f}_2 &=& x^4-12 x^3 z+90 x^2 z^2-324 x z^3+6 x+729 z^4-54 z , \cr
\tilde{f}_3 &=& 4 x \left(6 x^3+2 x^2 (2 y+z)-3 x z^2-2\right), \cr
\tilde{f}_4 &=& 2 \left(6 x^4+x^3 (4 y-34 z)-3 x^2 z (8 y-49 z)+2 x \left(54 y z^2+36 z^3-1\right)-81 z^4+6 z\right), \cr
\tilde{f}_5 &=& 4 x^2 \left(x^2-6 x z+27 z^2\right) .
\end{eqnarray}

We will extremize this potential to obtain the equations of motion. We need to use the method of 
Lagrange multipliers to incorporate the constraint \eqref{veq1m1}. We find 
\begin{eqnarray}
&&(p^1)^2 \left(-12 x^4 (2 y+z)-16 x^3 (y-z) (y+2 z)+18 x^2 z^2 (2 y+z)
+2 x \left(2 y-9 z^4+z\right)-3 z^2\right)\cr
&+&2 p^1 x \left(x^3 z (2 p^2-17 p^3)+x (2 p^2+p^3) \left(2 x^2 (3 x+y)-1\right)
-54 p^3 x y z^2-36 p^3 x z^3+81 p^3 z^4\right) \cr 
&+& 8 (p^2)^2 x^5+ p^3 x^2 \left(8 p^2 x^3-36 p^2 x z (x-3 z)
+p^3 \left(2 (x-3 z) \left(x^2-6 x z+27 z^2\right)+3\right)\right) = 0,\cr
&&(p^3 x-p^1 z) \Big(p^1 \left(2 x \left((x-9 z) \left(6 x^2+2 x (2 y+z)-3 z^2\right)-1\right)
+9 z\right)+4 p^2 x^3 (x-9 z) \cr
&+& p^3 x \left(2 (x-9 z) \left(x^2-6 x z+27 z^2\right)+9\right)\Big) = 0 \ .
\end{eqnarray}
To simplify these equations, we will use charge ratios defined as $r=p^1/p^3$ and $s=p^2/p^3$
and the inhomogeneous coordinates $\tau=x/z$ and $t=y/z$. In terms of these variables the above
equations read as 
\begin{eqnarray}
&& r^2 \left(4 \left(2 t^2+2 t-7\right) \tau ^3+8 (2 t+1) \tau ^4
-9 (2 t+1) \tau ^2-9 (4 t+1) \tau +27\right) \cr 
&-&2 r \tau  \left((8
s+4) \tau ^4+(6 s+3) \tau ^2-9 \tau  (2 s+6 t+5)-18 \tau ^3+81\right) \cr
&-&\tau ^2 \left(8 \left(s^2+s+1\right) \tau ^3-3 \tau ^2 (12 s-2 t+5)+27 (4 s+3)
\tau -135\right) = 0 \ ,\\
&& (r-\tau ) \Big(r \left(8 \tau ^4+4 (t-22) \tau ^3-27 (2 t+1) \tau ^2+9 \tau +81\right) 
+\tau  \big(4 (s+5) \tau ^3 \cr 
&-& 3 \tau ^2 (12 s-6 t+7)+135 \tau -405\big)\Big) = 0
\end{eqnarray}

We will now solve these equations to obtain black string configurations. It can easily be seen 
that $\tau=r, t=s$ solves these equations. This set of solutions corresponds to the BPS string.
To obtain the non-BPS black string solutions we will first consider the inverse problem where 
we express the charge rations $r$ and $s$ in terms of the inhomogeneous coordinates $\tau$ and $t$.
This can be easily obtained from the equations of motion. We find two independent solutions for
non-BPS black strings:
\begin{eqnarray} \label{1stset}
r = \tau , s = - \frac{1}{\tau^2} \big(2\tau^3 + \tau^2 (t+1) - 3\tau + 9\big) \ , 
\end{eqnarray}
and 
\begin{eqnarray}\label{nbpmr}
r = - \frac{\tau  \left(56 \tau ^3-(27-54 t) \tau ^2+405 \tau -1215\right)}{24 \tau ^4+4 (6 t-65) \tau ^3-9 (30 t+13) \tau ^2+27 \tau +243} \ , 
\end{eqnarray}
\begin{eqnarray}
s=- \frac{2 \tau^2 (4 \tau ^2- 34 \tau ^3 + 27 t^2 -24) +t \left(64 \tau ^3-99 \tau ^2+405 \tau -1215\right)+144 \tau -324}{24 \tau
^4+4 (6 t-65) \tau ^3-9 (30 t+13) \tau ^2+27 \tau +243} . \label{nbpms}
\end{eqnarray}
The first set of equations \eqref{1stset} can be easily solved for the scalar moduli to obtain 
\begin{equation} \label{fbr1}
\tau = r , t = - \frac{1}{r^2}\left(2 r^3 + r^2 (s+1) - 3 r + 9  \right)  \ . 
\end{equation}
Thus, for a given set of charges, we have a unique solution in this branch. The resulting solution 
will lie inside the K\"ahler cone if $r>3$ and $s<-(2r^3+r^2-3r+9)/r^2$.

We will now analyze the second branch of non-BPS solutions given in \eqref{nbpmr} and \eqref{nbpms}.
First solve \eqref{nbpmr} for $t$ to obtain
\begin{equation} \label{sbr1}
t = -\frac{r \left(24 \tau ^4-260 \tau ^3-117 \tau ^2+27 \tau +243\right)
+\tau  \left(56 \tau ^3-27 \tau ^2+405 \tau -1215\right)}{6\tau ^2 (r (4 \tau -45)+9 \tau )}
\end{equation}
Substituting this expression for $t$ in \eqref{nbpms} and simplifying we obtain 
\begin{eqnarray}\label{quartic}
&& 2 \tau ^4 \left(12 r^2+4 r (3 s+8)+27 s+9\right)-r \tau ^3 (268 r+270 s+135) 
- 27 (r-15) r \tau ^2 \cr &+& 27 (r-45) r \tau +243 r^2 = 0 \ . 
\end{eqnarray}
This is a quartic equation in $\tau$. Thus we can express the exact analytic solutions for $\tau$ 
in terms of the charge ratios $r$ and $s$. However, before we consider the exact solutions, we 
will qualitatively analyze the relevant equations for this branch of solutions.

From \eqref{nbpmr} and \eqref{nbpms} we find that the charge ratios $r$ and $s$ are expressed as 
rational functions of $\tau$ and $t$. The denominators of these rational functions vanish when
\begin{equation}
24 \tau^4+4 (6 t-65) \tau ^3-9 (30 t+13) \tau ^2+27 \tau +243 = 0 \ . 
\end{equation}
This defines a singularity curve in the moduli space. To see this, solve the above equation for $t$
to obtain
\begin{equation}
t = -\frac{24 \tau ^4-260 \tau ^3-117 \tau ^2+27 \tau +243}{6 \tau ^2 (4 \tau -45)}
\end{equation}
The numerator has a single zero at $\tau=\tau_0\simeq 11.2506$ in the $\tau>3$ region. The denominator
vanishes at $\tau=45/4$. The value of $t$ in the above equation remains positive for 
$45/4<\tau<\tau_0$. This gives rise to a line in the moduli space which starts at $\tau=\tau_0$ when
$t=0$ and asymptotes to $\tau=45/4$ line as $t\to\infty$. Points on this curve do not correspond
to any extremal black string configuration. This singularity curve divides the moduli 
space into two regions, say region 1 and region 2. As we approach it from either of 
the regions the charge ratios diverge. Any point interior to either of the regions 
corresponds to a possible non-BPS black string configuration. The equations of motion
\eqref{nbpmr} and \eqref{nbpms} guarantee the existence of such attractor configurations.
In region 1, at the boundary point $(\tau=3,t=0)$ the charge ratios take values 
$r=141/215$ and $s=-56/215$.
These are the minimum possible values for $r$ and $s$ in region 1.
These values change from point to point and they diverge as we 
approach the singularity curve from the left. Similarly we can 
notice that, in region 2 the charge ratios take the maximum values $r = -7/3$ and 
$s = -1/3$ as $\tau\to\infty$. For any finite $t$ these values decrease monotonically
with $\tau$ in this region and diverge as we go closer to the singularity curve from 
the right. 

The above analysis indicates that these solutions are mutually exclusive to the first 
branch of solutions given in \eqref{fbr1}. We would like to investigate whether there exist multiple black string configurations 
for the second branch of solution. We first consider extremizing $s$ in \eqref{nbpms} as a 
function of $\tau$ and $t$. Numerically solving the equations $\partial s/\partial\tau
= \partial s/\partial t=0$ we find that they do not admit any solution inside the 
K\"ahler cone. We can further check if there exists an extremum in $\tau$ for a fixed
value of $t$. Setting $\partial s/\partial\tau=0$ for fixed $t$, we find
\begin{eqnarray}
&& 108 \tau ^4 t^2 + 36  \tau  \left(6 \tau ^4+\tau ^3+45 \tau^2-459 \tau +1458\right) t
+\big(112 \tau ^6-108 \tau ^5+3561 \tau ^4-27522 \tau ^3 \cr
&+& 71685 \tau ^2+24786 \tau -10935\big) = 0 \ .
\end{eqnarray}
This is a quadratic equation in $t$ with positive coefficients in the $\tau>3$ region.
Thus, it does not admit any solution with $t>0$. As a result, $s$ does not admit an extremum
in $\tau$ for a fixed $t$ inside the K\"ahler cone. Hence there can't be any multiple 
solution for a given set of charges in this model.

We will now consider the exact solution of the quartic \eqref{quartic} for a given value of 
$r$ and $s$. Introducing the notation
\begin{eqnarray}
&& d_0 = 243 r^2 \ ,\; d_1 = 27 r (r-45) \ ,\; d_2 = - 27 r (r-15) \ , \cr 
&& d_3 = -r (268 r+270 s+135)\ ,\;
  d_4 = 2 \left(12 r^2+4 r (3 s+8)+27 s+9\right) \ ,
\end{eqnarray}
the equation \eqref{quartic} can be rewritten as 
$$\sum_{k=0}^4 d_k\tau^k=0 \ . $$
Depending on the sign of the coefficients $d_k$ this can admit zero, two or four real roots.
As we have argued above, for a given choice of the charge ratios at most one of these real 
roots will lie inside the K\"ahler cone. Denoting the solutions as $\tau_{1\pm}$ and $\tau_{2\pm}$
we have 
\begin{eqnarray*}
\tau_{1\pm} &=& - \frac{1}{12 d_4}\left(3 d_3 
+ \sqrt{3(9d_3^2 - 24 d_2 d_4 - 2{\mathcal{X}})}
\pm \sqrt{6 \mathcal{X}
+ \frac{18 \sqrt{3}(d_3^3 - 4 d_2 d_3 d_4 + 8 d_1 d_4^2)}
{\sqrt{3(9d_3^2 - 24 d_2 d_4 - 2{\mathcal{X}})}}}\right), \cr
\tau_{2\pm} &=& - \frac{1}{12 d_4}\left(3 d_3 
- \sqrt{3(9d_3^2 - 24 d_2 d_4 - 2{\mathcal{X}})}
\pm \sqrt{6 \mathcal{X}
+ \frac{18 \sqrt{3}(d_3^3 - 4 d_2 d_3 d_4 + 8 d_1 d_4^2)}
{\sqrt{3(9d_3^2 - 24 d_2 d_4 - 2{\mathcal{X}})}}}\right).
\end{eqnarray*}
Here, for easy reading we have introduced the notations
\begin{equation*}
e_1 = d_2^2  - 3 d_1 d_3 + 12 d_0 d_4 \ , \ 
e_2 = 2 d_2^3 - 9 d_2(d_1 d_3 + 8 d_0 d_4) + 27 (d_0d_3^2 + d_1^2 d_4) \ , 
\end{equation*}
in terms of which we define 
\begin{equation*}
e_3 = e_2 + \sqrt{e_2^2 - 4 e_1^3} \   {\rm and} \ 
\mathcal{X} = 3 d_3^2 - 8 d_2 d_4 - 2^{2/3} e_3^{1/3} d_4
- 2^{4/3} e_3^{-1/3} e_1 \ .
\end{equation*}

Though we obtained multiple black hole solutions in these three parameters THCY models, the 
black string solutions for a given set of magnetic charges are unique. While this might be 
true for the examples studied here this is not the case in general. To see this we will 
consider an explicit example of three parameter model admitting multiple black string 
solutions. The polytope ID associated with this specific THCY is 146. It has Hodge numbers
$h_{1,1}=3, h_{2,1}=81$ and Euler number $\chi=-156$. The resolved weight matrix for 
this manifold is 
\begin{equation}
Q = \left(
\begin{array}{ccccccc}
 0 & 0 & 1 & 1 & 1 & 0 & 1 \\
 0 & 1 & 0 & 1 & 0 & 1 & 1 \\
 1 & 0 & 0 & 1 & 0 & 0 & 1 \\
\end{array}
\right) \ .
\end{equation}
The non-vanishing intersection numbers associated with the THCY $\mathcal{M}$ are 
$f=1/2,g=4/3i=4/3,j=8$. Thus, the volume of $\mathcal{M}$ is given by
\begin{equation}
\mathcal{V} = 3 x y z+4 x z^2+4 y z^2+8 z^3 \ .
\end{equation}
The corresponding Mori cone matrix is 
\begin{equation}
\mathcal{M}^i_{~j} = \left(
\begin{array}{ccccccc}
 -1 & 0 & 1 & 0 & 1 & 0 & 0 \\
 -1 & 1 & 0 & 0 & 0 & 1 & 0 \\
 1 & 0 & 0 & 1 & 0 & 0 & 1 \\
\end{array}
\right) \ . 
\end{equation}
The divisor class $\{D_i\}$ is obtained from the $(1,1)$ homology
basis of $\mathcal{M}$ as $D_1 = J_3-J_1-J_2, D_2=J_2, D_3=J_1
D_4=J_3, D_5=J_1, D_6=J_2, D_7=J_3$.
From the expression for the Mori cone matrix, we find the K\"ahler cone matrix to be 
$3\times 3$ identity matrix. Thus the K\"ahler cone conditions are $x>0,y>0,z>0$ or,
in terms of the inhomogeneous coordinates $\tau>0, t>0$ and $z>0$. Using $\mathcal{V}=1$
constraint we can show that the condition $z>0$ can be derived from $\tau>0,t>0$ and 
need not be imposed separately.

We will study black string solutions in this model. The effective black string potential
is given by
\begin{eqnarray}
V = \frac{1}{3}\big((z (3 p^1 y+4 p^1 z+3 p^2 x+4 p^2 z)
+p^3 x (3 y+8 z)+8 p^3 z (y+3 z))^2 \cr
-2 (p^2 (3 p^1 z+3 p^3 x+8 p^3 z)+p^3 (p^1 (3 y+8 z)+4 p^3 (x+y+6 z)))\big) \ .
\end{eqnarray}
The equations of motion are given by
\begin{eqnarray}
&& (3 y + 4 z) (-p^3 (3 p^1 + 4 p^3) + (3 p^3 x + 3 p^1 z + 
       8 p^3 z) (8 p^3 z (y + 3 z) + p^3 x (3 y + 8 z) \cr 
    &&    + z (3 p^2 x + 3 p^1 y + 4 p^1 z + 4 p^2 z))) + (3 x + 
    4 z) (p^3 (3 p^2 + 4 p^3) - (3 p^3 y + 3 p^2 z + 
       8 p^3 z)  \cr 
       && (8 p^3 z (y + 3 z) + p^3 x (3 y + 8 z) + 
       z (3 p^2 x + 3 p^1 y + 4 p^1 z + 4 p^2 z))) = 0, \cr 
     &&  (8 z (y + 3 z) + 
    x (3 y + 8 z)) (p^3 (3 p^2 + 4 p^3) - (3 p^3 y + 3 p^2 z + 
       8 p^3 z) (8 p^3 z (y + 3 z) \cr 
    &&   + p^3 x (3 y + 8 z) + 
       z (3 p^2 x + 3 p^1 y + 4 p^1 z + 4 p^2 z))) + 
 z (3 y + 4 z) (-8 p^3 (p^2 + 3 p^3) \cr
&&  - 
    p^1 (3 p^2 + 8 p^3) + (8 p^3 (x + y + 6 z) + p^2 (3 x + 8 z) + 
       p^1 (3 y + 8 z)) (8 p^3 z (y + 3 z) \cr
    &&    + p^3 x (3 y + 8 z) + 
       z (3 p^2 x + 3 p^1 y + 4 p^1 z + 4 p^2 z))) = 0 \ .
\end{eqnarray}
We now express these equations in terms of the rescaled quantities to find 
\begin{eqnarray}
(r (3 t+4)-3 s \tau -4 s+4 t-4 \tau ) (r (3 t+4)+s (3 \tau +4)+4 (t+\tau +4))&=&0,\cr
r^2 (3 t+4)^2 (3 t+8)+8 r (3 t+4) \left(-4 s+3 t^2+18 t+16\right) && \cr
-4 s^2 \left(9 \tau ^2+42 \tau +40\right)-32 s \left(3 \tau ^2-3 (t-4) \tau +20\right)-27 \tau ^2 t^3-72 \tau  t^3 && \cr
-108 \tau ^2 t^2-360 \tau  t^2+160 t^2-128 \tau ^2-144 \tau ^2 t-448 \tau  t+640 t-512 \tau &=& 0 \ .
\end{eqnarray}
It is not instructive to express the analytical expressions for $\tau,t$ in terms of $r,s$. 
Instead, we will focus on the inverse problem and express the charge ratios as rational functions 
of the inhomogeneous coordinates. We find the BPS solution $r=\tau, s=t$ and three branches of 
non-BPS attractors. They are given by
\begin{eqnarray}
&& r = \tau \ , \ s =  -t-\frac{8 (\tau +2)}{3 \tau +4} \ , \label{m3s1}  \\
&& r = -3 \tau - \frac{8 (t+2)}{3 t+4} \ , \ s = t \ ,  \label{m3s2} \\
&&  r =  -\frac{3 \tau^2 (3 t +4)^2+32 \left(3 t^2+4 t -4\right)+36 \tau \left(3 t^2+10 t +8\right)}{(3 t +4) (9 \tau t +12(\tau+t) -40)} \ ,  \cr
&& s =  -\frac{3 t^2 (3 \tau +4)^2+32 \left(3 \tau ^2+4 \tau -4\right)+36 t \left(3 \tau ^2+10 \tau +8\right)}{(3 \tau +4) (9 \tau t +12(\tau+t) -40)} \ . \label{m3s3}
\end{eqnarray}

We can easily invert \eqref{m3s1} and \eqref{m3s2} to express the inhomogeneous coordinates in terms of 
$r,s$. We find
\begin{eqnarray}
&& \tau = r \ , \ t = -s - \frac{8 (r+2)}{3 r+4} \ , \label{s1m3} \\
&& \tau = - r - \frac{8 (s+2)}{3 s+4} \ , \ t = s \ . \label{s2m3}
\end{eqnarray}
The first solution \eqref{s1m3} lies within the K\"ahler cone for $r>0$ and $s<-8(r+2)/(3r+4)$ where
as the second solution \eqref{s2m3} lies within the K\"ahler cone for $s>0$ and $r<-8(s+2)/(3s+4)$.

To invert \eqref{m3s3}, we will first do a shift and rescale the quantities as 
$\tau=4(\tilde{\tau}-1)/3, t= 4(\tilde{t}-1)/3, r = 4(\tilde{r}-1)/3$ and $s = 4(\tilde{s}-1)/3$.
The equations take a simpler form in terms of these variables. We find 
\begin{equation}
\tilde{r} \tilde{t} (2 \tilde{t} \tilde{\tau}-7)+2 \tilde{t}^2 \tilde{\tau}^2+3 \tilde{t} \tilde{\tau}-3=0 \ , \ 
\tilde{s} \tilde{\tau} (2 \tilde{t} \tilde{\tau}-7)+2 \tilde{t}^2 \tilde{\tau}^2+3 \tilde{t} \tilde{\tau}-3=0 \ . \label{elmntu}
\end{equation}
Solving the above equations we find $\tilde{\tau}=\tilde{t}\tilde{r}/\tilde{s}$ with 
\begin{eqnarray}
\tilde{t}_{1\pm} &=& -\frac{1}{4}\left( \tilde{s}+ \sqrt{\tilde{s}^2 + 4 \tilde{g}}
\pm 2 \sqrt{\frac{\tilde{s}^2(\tilde{r}\tilde{s}-34)}{2\tilde{r}\sqrt{\tilde{s}^2 + 4 \tilde{g}}}+\frac{\tilde{r}\tilde{s}^2 - 2\tilde{g}\tilde{r}-6\tilde{s}}{2\tilde{r}}}\right) \ , \cr
\tilde{t}_{2\pm} &=& -\frac{1}{4}\left( \tilde{s}- \sqrt{\tilde{s}^2 + 4 \tilde{g}}
\pm 2 \sqrt{\frac{\tilde{s}^2(\tilde{r}\tilde{s}-34)}{2\tilde{r}\sqrt{\tilde{s}^2 + 4 \tilde{g}}}+\frac{\tilde{r}\tilde{s}^2 - 2\tilde{g}\tilde{r}-6\tilde{s}}{2\tilde{r}}}\right) \ . \label{4slns}
\end{eqnarray}
where $\tilde{g}$ is given by
\begin{equation}
\tilde{g} = \frac{\tilde{s}}{2\ 3^{2/3} \tilde{r}}
\left(\big(225(2\tilde{r}\tilde{s}+ 1) + \tilde{f}\big)^{1/3}
+\big(225(2\tilde{r}\tilde{s}+ 1) - \tilde{f}\big)^{1/3} 
- 2\ 3^{2/3}\right) \ .
\end{equation}
In the above we have used the notation
\begin{equation}
\tilde{f} = \sqrt{6(13068 + 24489\tilde{r}\tilde{s} + 39924 \tilde{r}^2 \tilde{s}^2 
- 1372 \tilde{r}^3 \tilde{s}^3)} \ .
\end{equation}

The solution will remain inside the K\"ahler cone for $\tilde{\tau}>1,\tilde{t}>1$. 
Though there are four solutions in \eqref{4slns} for a given value of $\tilde{r},\tilde{s}$,
all of those do not satisfy the K\"ahler cone condition simultaneously. We need to find how many of those solutions lie inside the K\"ahler cone. This of course will depend on the 
values of $\tilde{r},\tilde{s}$. Since $\tilde\tau= \tilde t\tilde r/\tilde s$ we can't 
have solutions satisfying the K\"ahler cone condition if $\tilde{r}$ and $\tilde{s}$ have 
opposite signs. In our notation, the K\"ahler cone condition for \eqref{s1m3} is $\tilde r
> 1, \tilde s<-1-1/\tilde r$, and for \eqref{s2m3} it is $\tilde{s}>1, \tilde r<-1-1/\tilde{s}$. Thus the three solutions \eqref{s1m3},\eqref{s2m3} and \eqref{4slns} are mutually
exclusive. Let us now consider the case when $\tilde r$ and $\tilde s$ have the same sign.
To understand this in detail let us eliminate $\tau$ from 
the equations of motion described by \eqref{elmntu} to obtain the following quartic in the variable $\tilde{t}$
\begin{equation}
 2\tilde r^2\tilde t^4 + 2\tilde r^2\tilde s\tilde t^3 + 3\tilde r\tilde s\tilde t^2 - 7\tilde r\tilde s^2\tilde t -3\tilde s^2 = 0 \ . \label{scng}
\end{equation}
First consider the case when both $\tilde r$ and $\tilde{s}$ are positive. In the quartic
\eqref{scng} the coefficients change sign only once. Thus, according to Descartes rule
of signs there is only one positive root for $\tilde{t}$. By suitable choice of the charge 
ratios we can make $\tilde{t}>1$ to obtain a unique non-BPS black string solution inside 
the K\"ahler cone. However, since the BPS solutions also exist for $\tilde{r}>1,\tilde{s}>1$ 
region, these two solutions are not mutually exclusive from each other. Finally consider the 
case where both $\tilde{r}$ and $\tilde{s}$ are negative. For charge ratios valued in this
range the coefficients in the quartic \eqref{scng} change three times. Thus the equation
can admit three positive roots for $\tilde t$. By suitable choice of the charge ratios it 
might  be possible to have more than one non-BPS attractors existing inside the K\"ahler 
cone. This is indeed the case as a numerical analysis shows. Here we numerically solve the 
equation \eqref{scng} for $\tilde{t}$ upon setting $\tilde{r}=-6n, \tilde{s}=-5n$ with 
$n=2,3,\ldots,10$. In each case there are three positive roots for $\tilde{t}$ with two 
of them having values greater than $1$. Since $\tilde{\tau}=\tilde{t}\tilde{r}/\tilde{s}
>\tilde{t}$ for these choices of charge ratios, we have multiple non-BPS black strings.
In Fig.\ref{m3p3} we have presented the two black string solutions obtained for different
values of $n$.
\begin{center}
\includegraphics[width=.49\linewidth]{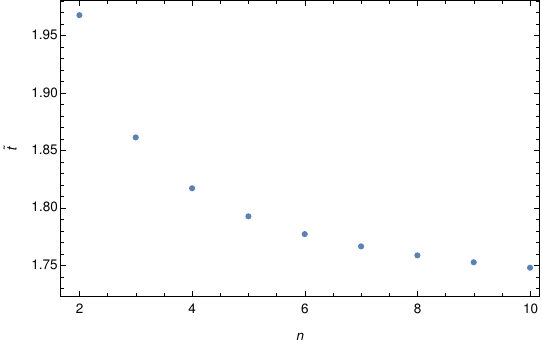}
\includegraphics[width=.49\linewidth]{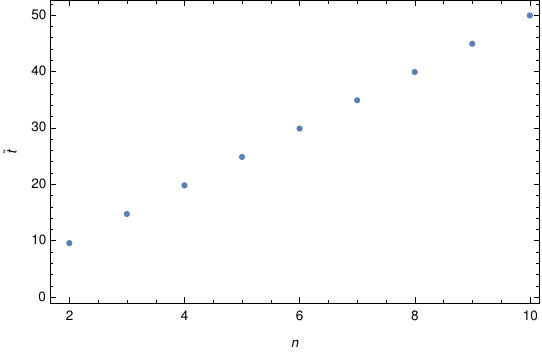}
\captionof{figure}{Multiple non-BPS black strings in model 3 for $\tilde{r}=-6n, \tilde{s}
= -5n$ with $n=2,3,\ldots,10$. In the first case the value of $\tilde{t}$ decreases with
increasing $n$ whereas in the second case it increases with $n$.}
\label{m3p3}
\end{center}

\subsection{Stability} \label{stbl2}

We will now analyze stability of these non-BPS black strings. For this we need to compute the 
recombination factor for the black string. The recombination factor is defined as the ratio
of the volume of the non-holomorphic divisor on which the $M5$ brane wraps to give rise to the
non-BPS black string, to the volume of minimal piece-wise calibrated cycle in the same 
homology class as the non-holomorphic divisor under consideration. For double extremal black 
strings, the asymptotic volume of the non-holomorphic divisor is given by the square root 
of the effective black string potential at the attractor value. For a black string of charge
$p^I$, the volume of the minimal piece-wise calibrated cycle is $\sum_I|p^I|C_I$,
where $C_I$ is the volume of the divisor $J_I$:
\begin{equation}
C_I = \int_{J_I} J\wedge J = C_{IJK} t^J t^K \ . 
\end{equation}
Denoting $R$ as the recombination factor, we have 
\begin{equation}
R = \frac{\sqrt{V_{\rm cr}}}{\sum_I |p^I|C_I} \ . 
\end{equation}

For a three parameter model, the recombination factor is given as 
\begin{equation} \label{strcmb}
R = \frac{\sqrt{V_{\rm cr}}}{|p^1| C_1 + |p^2| C_2 + |p^3| C_3} \ , 
\end{equation}
where the volumes $C_I$ are given in \eqref{cijtj}. We can substitute the expression for
the black string solution in it to obtain the value of $R$. We will first compute the 
recombination factor for non-BPS black strings in model 1 before turning our discussion 
into the stability issues solutions in model 2. In the case of model 1, there are three
branches of non-BPS solutions. Let us first consider the recombination factor for the 
first branch of solutions given in \eqref{br1st1}. In this case $s>0$ and $r<2-4s$. We will consider the cases $0<r<2-4s$ and $r<0$ separately.
Substituting the explicit expressions for the solution \eqref{br1st1} in
\eqref{strcmb} and simplifying, we find the recombination factor in the 
case $s>0$ and $0<r<2-4s$  to have the form 
\begin{equation}
R = 3 \left(2 s(s+1) (2-r-2s)\right)^{1/3} \ .
\end{equation}
In this case, $R$ always remains greater than one if $s>s_0$, where $s_0\simeq 0.09$ is a root of 
the equation $108(s^3+s^2)=1$ and the resulting black strings are all unstable. For $0<s<s_0$ the 
black strings are stable if ${\rm max}\{0,-(108s^3-108s+1)/54s(s+1)\}<r<2-4s$ and unstable otherwise.

For $s>0$ and $r<{\rm min}\{0,2-4s\}$ the recombination factor becomes
\begin{equation}
R = \frac{3 \left(2 s(s+1)\right)^{1/3} (2-r-2s)^{4/3}}{(2-3r-2s)} \ .
\end{equation}
In this case $R>1$ for in the entire range of allowed values for $r$ and $s$. Thus, in this case 
the black strings are all unstable.

For the second branch \eqref{br2st1}, $r$ and $s$ are valued in the range $-1/2<s<0, r>2-4s$. 
The recombination factor for this branch is given by
\begin{equation}\label{rro}
R = \frac{3  (-2s)^{1/3}\left((s+1)(2s+r-2)\right)^{4/3}}{(2s+1)\big(r(s+3)-2(s^2+3)\big)}
\end{equation}
From the above expression we find that there is no stable attractor for $-1/2<s<-1/3$ irrespective
of the values taken by $r$. For $-1/3<s<0$, the attractors become stable if $2-4s<r<r_0$, where the 
value of $r_0$ is obtained by solving $R=1$ in \eqref{rro} for $r$ for a fixed value of $s$.

Finally we consider the third branch of solutions \eqref{br3st1}. In this case $-1/2<s<0$ and $r<2$. We will first
consider the range $-1/2<s<0, 0<r<2$. In this case the recombination factor is
\begin{equation}
R = \frac{3 (-2s)^{1/3} \left((s+1)(r+2s-2)\right)^{4/3}}{(2s+1)(2s^2 - rs+r+8s-2)} \ . 
\end{equation}
In this case the black strings are unstable for $-1/2<s<s_0$, where $s_0\simeq - 0.26$ is a root of
$108 s^2 (s+1)^4 - (s+3)^3 (2 s + 1)^3=0$. For $s_0<s<0$ the black strings become unstable in the 
range $0<r<r_0$ and stable for $r_0<r<2$, where the value of $r_0$ for a fixed $s$ is determined by
$54 s (s+1)^4 (r+2 s-2)^4-(2 s+1)^3 (2 s^2-r s+r+8 s-2)^3=0$. 

Consider now the third branch of solutions \eqref{br3st1} with $-1/2<s<0$ and $r<0$. In this case 
the recombination factor becomes
\begin{equation} \label{rr1}
R = \frac{3 (-2s)^{1/3} \left((s+1)(r+2s-2)\right)^{4/3}}{(2s+1)\left(r(s+3)+2(s^2+4s-1)\right)} \ . 
\end{equation}
In this case the attractors are unstable for $-1/2<s<s_0$ where $s_0\simeq = - 0.036$ is a root of 
the equation $2048 s (s+1)^5+(s+3)^4 (2 s+1)^3=0$. For $s_0<s<0$ the attractors are stable for 
$r_2<r<r_1$ where $r_{1,2}$ are obtained by solving $R=1$ in \eqref{rr1} for $r$ for a fixed value 
of $s$.

Let us focus on the recombination factor for non-BPS black strings in model 2. We will 
first consider the solution \eqref{fbr1}. In this 
case, the recombination factor \eqref{strcmb} takes the simple form
\begin{equation}
R = \frac{3 r\big(2r^2 + 2 r s + r - 3\big) + 27}{r\big(2r^2 + 6 r s + r - 3\big) + 9} \ . 
\end{equation}
Recall that the solution \eqref{fbr1} exists for $r>3$ and $s<-(2 r^3 + r^2 - 3 r +9)/r^2$.
For any given $r$, the recombination factor takes the minimum value $R=3/5$ when $s$ takes
the maximum value $s=-(2 r^3 + r^2 - 3 r +9)/r^2$. The value of $R$ increases as $s$ becomes
more negative and approaches the value $R=1$ as $s\to -\infty$. Thus $R<1$ for all the black 
strings in this branch of solutions and hence they are all stable.

We will now consider the recombination factor for the second branch of solutions described 
by \eqref{sbr1} and \eqref{quartic}. The resulting expression is complicated and it is not 
possible to extract any insight from it. In what follows, we will numerically solve the
equations of motion for various choices of the charge ratios and substitute the resulting
solutions to obtain the values of the respective recombination factors. We first consider  
a fixed value for $s, (s=3/2)$ and vary $r$ in the range $1\leq r\leq 5/2$. This leads 
solutions in region 1. We notice the the value of $\tau$ increases monotonically with the
increase of $r$ while the value of $t$ decreases. The plot for the inhomogeneous coordinates 
$\tau$ and $t$ are given in Fig.\ref{mgtr1}. Further we consider the value $s=-3/2$ and vary 
$r$ in the range $-24\leq r\leq -3$ to obtain solutions in region 2. In this case, both the
inhomogeneous coordinates increase with $r$. The results are given in the Fig.\ref{mgtr2}.
\begin{center}
\includegraphics[width=0.49\linewidth]{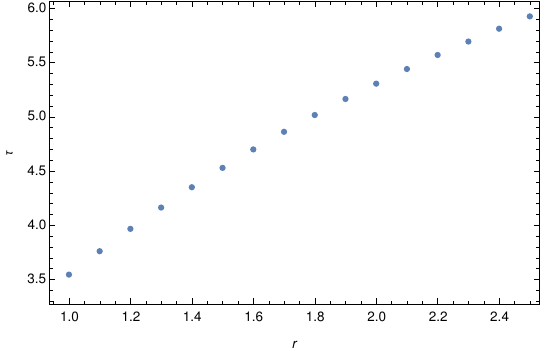}
\includegraphics[width=0.49\linewidth]{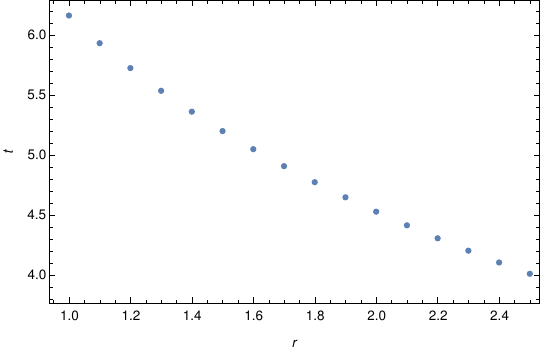}
\captionof{figure}{Non-BPS black string solutions in region 1 with $s=3/2$ and $1\leq r\leq 5/2$.} \label{mgtr1}
\end{center}
\begin{center}
\includegraphics[width=0.49\linewidth]{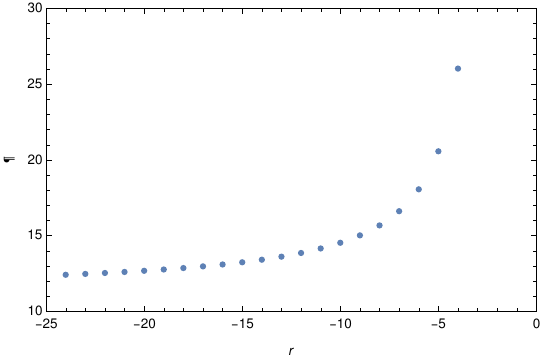}
\includegraphics[width=0.49\linewidth]{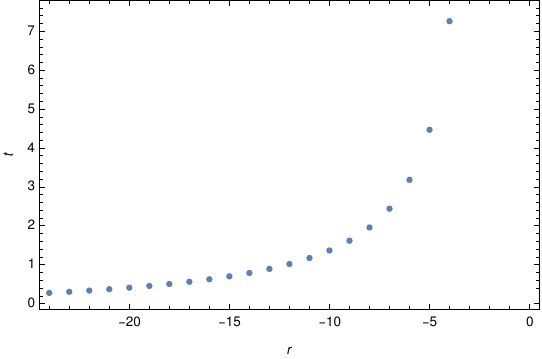}
\captionof{figure}{Non-BPS black string solutions in region 2 with $s=-3/2$ and $-24\leq r\leq -3$.} \label{mgtr2}
\end{center}

The recombination factors for solutions in both the regions are depicted in Fig.\ref{mgtr3}.
In region 1 the value of recombination factor increases with $r$ whereas in region 2 it 
decreases. We observe that the value of $R$ remains less than one in both the regions and it 
tends to saturate to $1$ as we approach the singularity curve. Thus the corresponding black 
strings in the interior of regions 1 and 2 are all stable. We have considered a wide range of 
charges and numerically evaluated the recombination factor for all these choices. We could 
not find any case with $R>1$. This suggests that the black strings corresponding to the second 
branch of solutions are also stable.

\begin{center}
\includegraphics[width=0.49\linewidth]{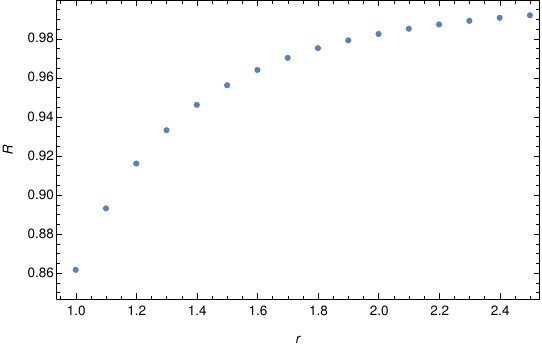}
\includegraphics[width=0.49\linewidth]{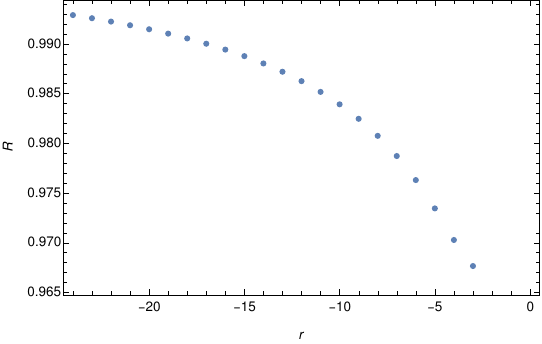}
\captionof{figure}{Recombination factors corresponding to the solutions in both the regions. The 
figure in the left corresponds to recombination factors in region 1 and the one in the right
corresponds to recombination factors in region 2.} \label{mgtr3}
\end{center}

Finally we turn our attention to the stability of non-BPS black strings in model 3. We will first
consider the solutions given in \eqref{s1m3} and \eqref{s2m3}. The recombination factors for these
two cases are given respectively as 
\begin{eqnarray}
R = \frac{3 (3 r s+4 r+4 s+8)}{9 r s+4 r+12 s+8}  \ {\rm and} \ 
R = \frac{3 (3 r s+4 r+4 s+8)}{9 r s+12 r+4 s+8} \ .
\end{eqnarray}
In both these cases the value of $R$ remains less than $1$ in the allowed range of the charge 
ratios. We can check this easily by expressing $R$ in terms of the inhomogeneous coordinates. 
In both the cases, $R$ has a minimum value of $3/5$ and it increases monotonically with $\tau$
and $t$ and approaches the maximum value $R=1$ when either of these coordinates becomes infinite.
Thus the corresponding black strings are all stable. For the solutions \eqref{scng} we 
have obtained the recombination factors numerically. As we can see from Fig.\ref{mm3p5}, depending
the value of $n$ the recombination factor becomes larger or smaller than one. Thus, in this case
we have both stable as well as unstable non-BPS black string solutions.
\begin{center}
\includegraphics[width=.49\textwidth]{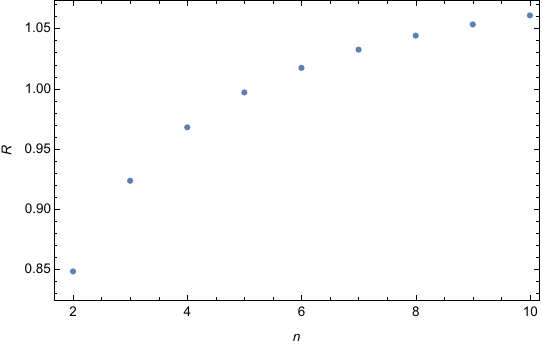}
\includegraphics[width=.49\textwidth]{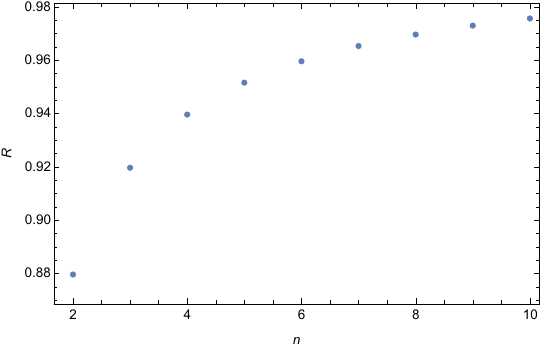}
\captionof{figure}{Recombination factors for the multiple solutions corresponding to Fig.\ref{m3p3}.}\label{mm3p5}
\end{center}

\section{Conclusion} \label{cncl}

In this paper we have studied BPS as well as non-BPS black branes in five dimensional supergravity
theories arising from the compactification of M-theory on three parameter Calabi-Yau manifolds.
We considered two explicit examples of toric Calabi-Yau manifolds with $h_{1,1}=3$ and obtained all
BPS and non-BPS black hole configurations in them. For the first model, the resulting BPS solutions 
were unique for a given set of charges. However, for model 2, we found multiple BPS black hole
solutions when their charges are valued in a particular range. The non-BPS black holes for model 1
admit three independent solutions. However, they are mutually exclusive from each other and for a 
given set of black hole charges, there is a unique solution. For the non-BPS black holes in model 2,
the moduli space admits two singularity curves. Points on the singularity curves do not correspond 
to any black hole solution. They divide the moduli space into three regions. Any point in a given
region corresponds to a possible non-BPS attractor for some suitable choice of black hole charges.
As we change the charges the points move within a given region. But they never cross the singularity
curves for finite values of black hole charges. This model also admits multiple non-BPS black holes.
We considered the stability of doubly extremal non-BPS black hole configurations in these models and 
found that the black holes become stable for a range of charges. This is in contrast to the examples
studied in \cite{Long:2021lon} and the two parameter models in \cite{Marrani:2022jpt} where the 
non-BPS black hole configurations were all unstable. Moreover none of the models considered in
\cite{Long:2021lon,Marrani:2022jpt} admitted multiple black hole solutions.

Subsequently we have studied non-BPS black string solutions in these models. Once again we found 
three independent non-BPS solutions in model 1. They were all mutually exclusive and hence we have
unique non-BPS black string solution for a given set of charges. Model 2 admitted two branches of
solutions. The second branch of solutions admitted a singularity curve which divided the moduli
space into two regions. Points on the singularity curve did not correspond to non-BPS attractors 
for any choice of the black string charges. Further, we found that for a given set of charges the 
resulting black string solution was always unique. To demonstrate the existence of multiple non-BPS
black string solutions we considered one more three parameter Calabi-Yau model. We obtained all
black string solutions in this model and observed that it admits multiple non-BPS solutions in
a given range of black string charges. We have also analyzed the stability of the doubly extremal 
black string configurations corresponding to all the attractor values we have obtained. It is 
interesting to observe so much rich structure for attractor configurations in three parameter 
Calabi-Yau models. It would be worth investigating the behaviour of both BPS as well as non-BPS 
multiple attractor configurations in more detail. In four dimensions, incorporating $D6$ 
branes gives a rich structure both in the BPS as well as non-BPS sector \cite{Tripathy:2005qp,Mandal:2015mke,Marrani:2017ivp,Tripathy:2017pwi}. Extending the analysis of \cite{Aspman:2022vlx} to 
incorporate the $D6$ branes and relating them with the solutions studied here using the 4D-5D
correspondence \cite{Ceresole:2007rq} is another interesting aspect that needs to be explored. 
We hope to analyze these issues in future.

\section*{Acknowledgments} 

We would like to thank Alessio Marrani for collaboration in \cite{Marrani:2022jpt} as well as for 
useful correspondences.


\end{document}